\documentclass[namedreferences]{solarphysics}
\usepackage[hyperref,optionalrh]{spr-sola-addons}
\usepackage{graphicx}
\usepackage{amssymb}            
\usepackage{color}                                        
\usepackage{breakurl}

\newcommand{\arcsec}{^{\prime\prime}}

\chardef\us=`\_

\begin{document}

\begin{article}
\begin{opening}
 
\title{Rapid Rotation of an Erupting Prominence and the Associated Coronal Mass Ejection on 13 May 2013}

\author[addressref={aff1,aff2},corref]{\inits{Y.H. Zhou}\fnm{Yuhao}~\lnm{Zhou}}
\author[addressref={aff1},corref]{\inits{H.S. Ji}\fnm{Haisheng}~\lnm{Ji}\orcid{0000-0002-5898-2284}}
\author[addressref={aff1},corref,email={zhangqm@pmo.ac.cn}]{\inits{Qingmin}\fnm{Qingmin}~\lnm{Zhang}\orcid{0000-0003-4078-2265}}

\address[id=aff1]{Key Laboratory of Dark Matter and Space Astronomy, Purple Mountain Observatory, CAS, Nanjing 210023, China}
\address[id=aff2]{School of Astronomy and Space Science, University of Science and Technology of China, Hefei 230026, China}

\runningauthor{Zhou et al.}
\runningtitle{Rapid rotation of an erupting prominence and the associated CME}

\begin{abstract}
In this paper, we report the multiwavelength observations of an erupting prominence and the associated coronal mass ejection (CME) on 13 May 2013.
The event occurs behind the western limb in the field of view of the Atmospheric Imaging Assembly (AIA) on board the Solar Dynamics Observatory (SDO) spacecraft.
The prominence is supported by a highly twisted magnetic flux rope and shows rapid rotation in the counterclockwise direction during the rising motion.
The rotation of the prominence lasts for $\sim$47 minutes.
The average period, angular speed, and linear speed are $\sim$806 s, $\sim$0.46 rad min$^{-1}$, and $\sim$355 km s$^{-1}$, respectively.
The total twist angle reaches $\sim$7$\pi$, which is considerably larger than the threshold for kink instability.
Writhing motion during 17:42$-$17:46 UT is clearly observed by SWAP in 174 {\AA} and 
Extreme-UltraViolet Imager (EUVI) on board the behind Solar Terrestrial Relations Observatory (STEREO) spacecraft in 304 {\AA} after reaching an apparent height of $\sim$405\,Mm.
Therefore, the prominence eruption is most probably triggered by kink instability. A pair of conjugate flare ribbons and post-flare loops are created and observed by STA/EUVI. 
The onset time of writhing motion is consistent with the commencement of the impulsive phase of the related flare.
The 3D morphology and positions of the associated CME are derived using the graduated cylindrical shell (GCS) modeling.
The kinetic evolution of the reconstructed CME is divided into a slow-rise phase ($\sim$330 km s$^{-1}$) and a fast-rise phase ($\sim$1005 km s$^{-1}$) by the writhing motion.
The edge-on angular width of the CME is a constant (60$^{\circ}$), while the face-on angular width increases from 96$^{\circ}$ to 114$^{\circ}$, indicating a lateral expansion.
The latitude of the CME source region decreases slightly from $\sim$18$^{\circ}$ to $\sim$13$^{\circ}$, implying an equatorward deflection during propagation.
\end{abstract}
\end{opening}

\section{Introduction}  \label{s-Intro}
Prominences are cool and dense plasmas suspended in the solar atmosphere \citep{Lab2010,Mac2010,Par2014,Chen2020}.
The materials are formed at magnetic dips due to levitation from below the photosphere \citep{Lit2005}, direct mass injection \citep{Wang2019}, and
catastrophic condensation as a result of thermal instability of evaporated hot plasmas from the chromosphere \citep{Xia2011,Luna2012,Zhou2014,Guo2021}.
Two types of magnetic configurations are believed to provide upward tension force to balance the gravity of prominences: sheared arcades \citep{Kip1957,Zha2015} 
and twisted flux ropes \citep{Kup1974,Low1995,Shi1995,Rust1996,Guo2022}.
\citet{Liu2012} proposed the scenario of double-decker structures where one flux rope is on top of another flux rope \citep{Hou2018} or a sheared arcade \citep{Awa2019}.
Both of them share a common polarity inversion line (PIL).  
According to their locations on the solar disk, prominences are divided into quiescent, active region, and intermediate types.
Prominences appear not only in H$\alpha$ \citep{Chae2008,Zha2022a} and Ca\,{\sc ii}\,H \citep{Ber2008,Zha2012} wavelengths, 
but also in ultraviolet (UV) and extreme-ultraviolet (EUV) wavelengths \citep{Zou2019}.
Prominences are also named filaments consisting of ultrafine dark threads on the disk \citep{Wang2015,Wang2018}.

Eruptions of small-scale filaments (minifilaments) can lead to blowout coronal jets \citep{Moo2010,Hong2013,Li2015,Ste2015,Zha2016,Wyp2018,Pan2022}.
However, eruptions of large-scale filaments are capable of producing solar flares and coronal mass ejections \citep[CMEs;][]{Chen2011,Geo2019},
which are frequently accompanied by EUV waves and coronal dimmings \citep{Thom1998,Shen2012,Zha2022c}.
The triggering mechanisms of filament eruptions include magnetic flux emergence \citep{Chen2000}, catastrophic loss of equilibrium of a magnetic flux rope \citep{Lin2000},
tether-cutting of magnetic sheared arcades \citep{Moo2001,Xue2017}, breakout model \citep{Anti1999,Chen2016}, ideal kink instability \citep{Kli2004,Tor2004,Fan2003,Fan2005,Tor2005}, 
and torus instability \citep{Kli2006}. The total twist of a helical flux tube is expressed as \citep{Hood1981,Shen2011b}:
\begin{equation} \label{eqn-1}
\Phi=\frac{LB_{\phi}(r)}{rB_{z}(r)},
\end{equation}
where $L$ and $r$ are the total length and radius of the tube, $B_{\phi}(r)$ and $B_{z}(r)$ denote the azimuthal and axial components of the magnetic field.
The number of turns over the tube length is \citep{Sri2010,Liu2016,Jing2018}:
\begin{equation} \label{eqn-2}
\mathcal{T}_w=\frac{1}{4\pi}\int_{L}\frac{\nabla\times \mathbf{B}\cdot \mathbf{B}}{B^2}dl=\frac{\Phi}{2\pi}.
\end{equation}
The critical values of $\Phi_c$ for kink instability ranges from 2.5$\pi$ to 3.5$\pi$ under different circumstances \citep{Hood1981,Tor2003,Tor2004,Kli2012}.
Observational evidences of helical kink instability are substantial in failed eruptions \citep[e.g.,][]{Ji2003,Alex2006,Liu2007,Liu2009,Guo2010,Ama2018}, 
partial eruptions \citep{Tri2013}, and successful eruptions \citep[e.g.,][]{Wil2005,Cho2009,Kum2012,Cheng2014a,Vem2017}.
The total twist is reported to be between 3.6$\pi$ and 12$\pi$, part of which is converted into writhe during eruption \citep{Gil2007,Tor2010}.
The sense of twist should be the same as the writhe \citep{Rust2005}.
Therefore, rapid rotation of filament legs or spines is frequently observed \citep{Gre2007,Liu2007,Bem2011,Kli2012,Su2013,Yan2014b}.

As mentioned above, filament eruptions could drive CMEs. The kinematic and morphological evolutions of CMEs in the three-dimensional (3D) space 
is essential for a precise determination of the arrival times of halo CMEs \citep{Liu2010}. 
To this end, various kinds of cone models were proposed assuming an ice-cream shape and a constant speed of a CME \citep[e.g.,][]{How1982,Zhao2002,Mich2003,Xie2004,Xue2005}.
These models have achieved great success in reconstructing the 3D morphology and tracking the propagation of earthward CMEs in the corona and interplanetary space.
Recently, \citet{Zha2021} put forward a revised cone model to explain non-radial prominence eruptions. 
The model is gratifyingly applied to tracking the 3D evolution of a halo CME due to the non-radial eruption of a flux rope on 21 June 2011 \citep{Zha2022b}.
Considering the flux-rope nature of CMEs \citep{Cre2004,Kra2006}, a new technique was developed to model the flux-rope like CMEs by using the graduated cylindrical shell (GCS) model \citep{The2006,The2009,The2011}.
This model is mainly characterized by three geometric parameters ($\alpha$, $h$, and $\kappa$) and three positioning parameters ($\phi$, $\theta$, and $\gamma$), 
which will be described in detail in Sect.~\ref{s-gcs}. 
This technique has widely been used to investigate the geometrical and kinematic evolutions of CMEs 
viewed from two or three perspectives \citep{Mie2009,Liu2018,Cre2020,Gou2020,Maj2020,Maj2022}.

\citet{Cheng2013} studied the eruptions of two successive magnetic flux ropes as a result of ideal torus instability 
and performed 3D reconstructions of the associated CMEs using GCS modeling.
Furthermore, \citet{Cheng2014b} tracked the evolution of a magnetic flux rope from the inner to the outer corona 
and found that the impulsive acceleration phase of the flux rope is caused by the torus instability. The 3D morphology of the associated CME was also obtained using GCS modeling.
So far, 3D reconstructions of CMEs caused by the eruption of flux ropes due to kink instability have rarely been reported.
In this paper, we carry out a detailed investigation of the rapid rotation of an erupting quiescent prominence and the associated CME occurring in the northern hemisphere on 13 May 2013. 
The prominence originates from the farside near the western limb, and the rotation is found to result from kink instability.
Meanwhile, we carry out a 3D reconstruction of the associated CME using the GCS modeling. The paper is organized as follows.
We describe the data analysis in Section~\ref{s-Data}. The results are presented in Section~\ref{s-Res} and compared with previous findings in Section~\ref{s-Dis}.
Finally, a brief summary is given in Section~\ref{s-Sum}.

\section{Data Analysis} \label{s-Data}
The prominence eruption was observed by the Atmospheric Imaging Assembly \citep[AIA;][]{Lem2012} on board the Solar Dynamics Observatory \citep[SDO;][]{Pes2012} spacecraft.
SDO/AIA takes full-disk images in seven EUV wavelengths (94, 131, 171, 193, 211, 304, 335 {\AA}) and two UV wavelengths (1600 and 1700 {\AA}).
The AIA level\_1 data were calibrated using the standard Solar Software (SSW) program \textsf{aia\_prep.pro}.
The prominence was also observed in 174 {\AA} ($\log T\approx5.8$) by the Sun Watcher using Active Pixel System detector and image processing \citep[SWAP;][]{Ber2006,Sea2013}
on board the PROBA 2 spacecraft with a larger field of view (FOV) than AIA.

\begin{figure}
\centerline{\includegraphics[width=0.6\textwidth,clip=]{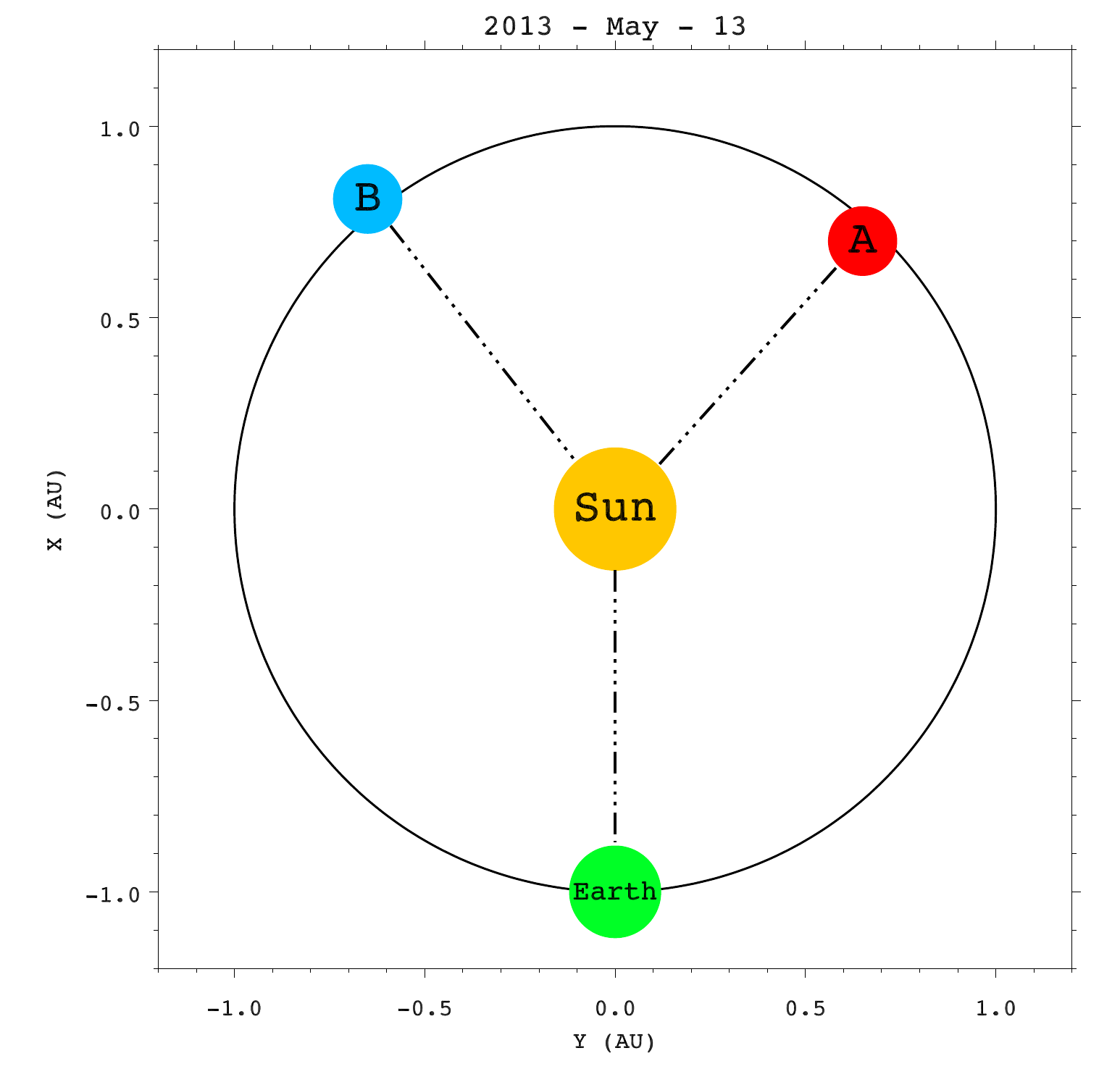}}
\caption{Positions of the Earth, Ahead (A), and Behind (B) STEREO spacecrafts at 00:55 UT on 13 May 2013.}
\label{fig1}
\end{figure}

The event was simultaneously captured by the ahead and behind Solar Terrestrial Relations Observatory \citep[STEREO;][]{Kai2008} spacecraft from two perspectives.
Figure~\ref{fig1} shows the positions of the Earth, Ahead (A), and Behind (B) STEREO spacecraft on 13 May 2013. 
The separation angles of STEREO-A and STEREO-B with the Earth were $\sim$136$^{\circ}$ and $\sim$142$^{\circ}$ \citep{Gou2020}.
Consequently, the large prominence close to the western limb in the FOV of AIA appeared as a prominence close to the eastern limb in the FOV of 
behind Extreme-UltraViolet Imager \citep[EUVI;][]{Wue2004} 
of the Sun Earth Connection Coronal and Heliospheric Investigation \citep[SECCHI;][]{How2008} and appeared as a long filament in the FOV of ahead EUVI.
This provides a unique opportunity to study the prominence eruption from three perspectives simultaneously \citep{Cheng2014b}.
EUVI takes full-disk images out to 1.7\,$R_{\odot}$ in 171, 195, 284, and 304 {\AA}. The images were calibrated using the SSW program \textsf{secchi\_prep.pro}.

The associated CME was observed by the Large Angle Spectroscopic Coronagraph \citep[LASCO;][]{Bru1995} on board SOHO spacecraft
and recorded by the CDAW CME catalogue\footnote{http://cdaw.gsfc.nasa.gov/CME\_list}.
The LASCO-C2 and LASCO-C3 white light (WL) coronagraphs have FOVs of 2$-$6\,$R_{\odot}$ and 4$-$30\,$R_{\odot}$, respectively.
The CME was also detected by the COR1 and COR2 coronagraphs on board STEREO-A and STEREO-B, which enables 3D reconstruction using the GCS modeling.
Soft X-ray (SXR) light curves of the Sun in 1$-$8 {\AA} and 0.5$-$4 {\AA} were recorded by the Geostationary Operational Environmental Satellite (GOES) spacecraft.
The observational parameters are listed in Table~\ref{tab-1}.

\begin{table}
\caption{Description of the observational parameters.}
\label{tab-1}
\tabcolsep 1.5mm
\begin{tabular}{lcccc}
  \hline
Instrument & $\lambda$   & Time &  Cadence & Pixel Size \\ 
                  & [{\AA}]         &  [UT] &  [s]           & [$\arcsec$] \\
  \hline
AIA & 171, 304 & 13:30\,--\,18:30 &  12 & 0.6 \\
SWAP       & 174       & 17:09\,--\,17:51 & 130 & 3.2 \\
STA/EUVI & 195       & 13:30\,--\,20:30 & 300 & 1.6 \\
STA/EUVI & 304       & 13:30\,--\,20:30 & 600 & 1.6 \\
STA/EUVI & 284       & 13:30\,--\,20:30 & 7200 & 1.6 \\
STB/EUVI & 304       & 13:30\,--\,20:30 & 600 & 1.6 \\
COR1 & WL      & 17:15\,--\,18:20 & 900 & 14.7 \\
COR2 & WL      & 17:30\,--\,20:00 & 900 & 14.7 \\
LASCO/C2 & WL & 17:24\,--\,19:36 & 720 & 11.4 \\
LASCO/C3 & WL & 18:18\,--\,19:54 & 720 & 56.0 \\
GOES     & 0.5\,--\,4 & 15:00\,--\,18:00 &  2.05 & ... \\
GOES     & 1\,--\,8      & 15:00\,--\,18:00  &  2.05 & ... \\
  \hline
\end{tabular}
\end{table}

\section{Results} \label{s-Res}

\subsection{Prominence Eruption and the associated CME} \label{s-prom}
\begin{figure} 
\centerline{\includegraphics[width=0.9\textwidth,clip=]{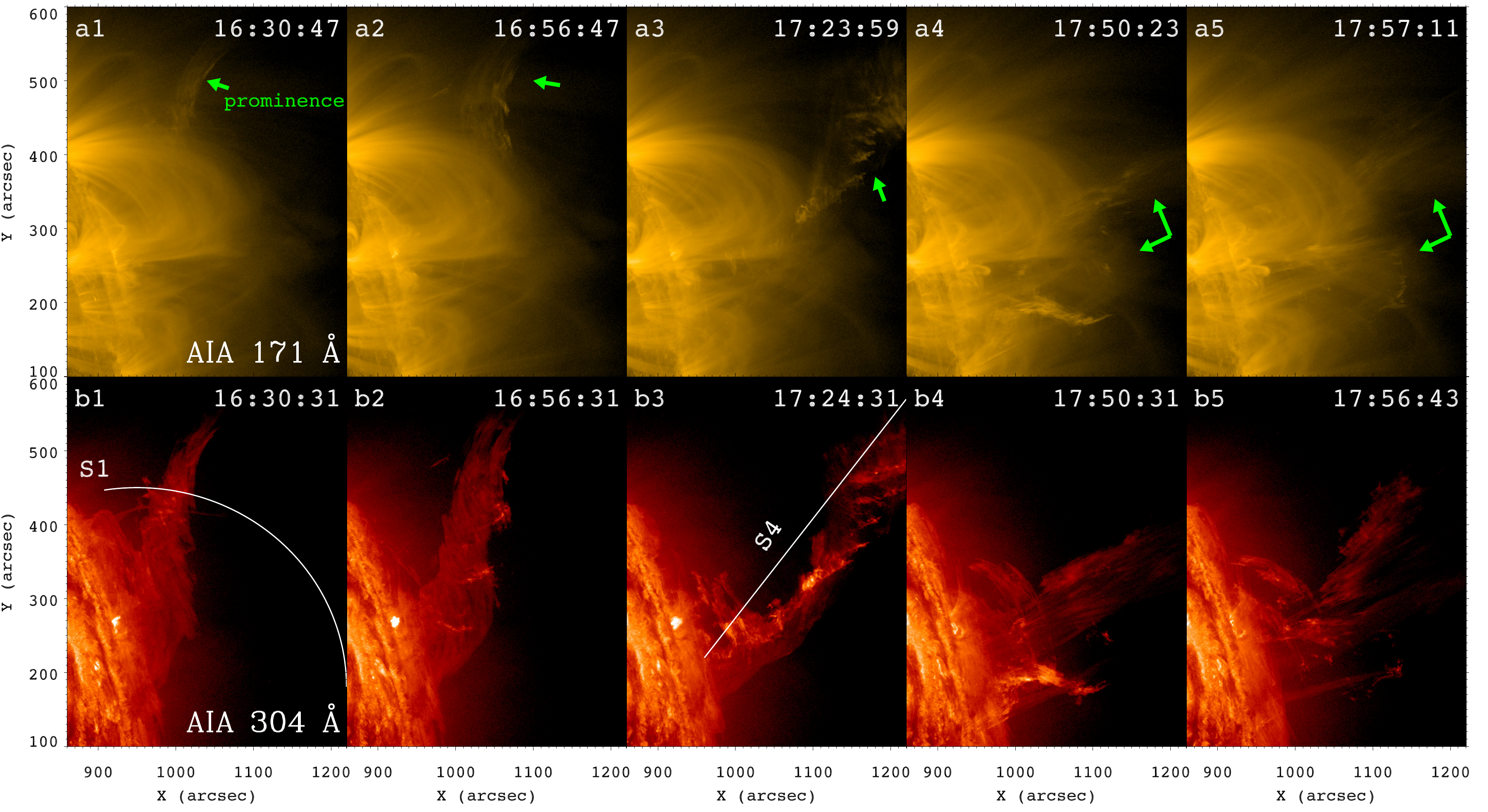}}
\caption{Selected 171 {\AA} (top panels) and 304 {\AA} (bottom panels) images observed by SDO/AIA during the prominence eruption, respectively.
In panel (b1), the curved slice S1 is used to investigate the rapid rotation and drift motion. In panel (b3), the straight slice S4 is used to investigate the rising motion.
An animation of this figure is available in the Electronic Supplementary Material (\textsf{anim1.mp4}).}
\label{fig2}
\end{figure}

\begin{figure}
\centerline{\includegraphics[width=0.9\textwidth,clip=]{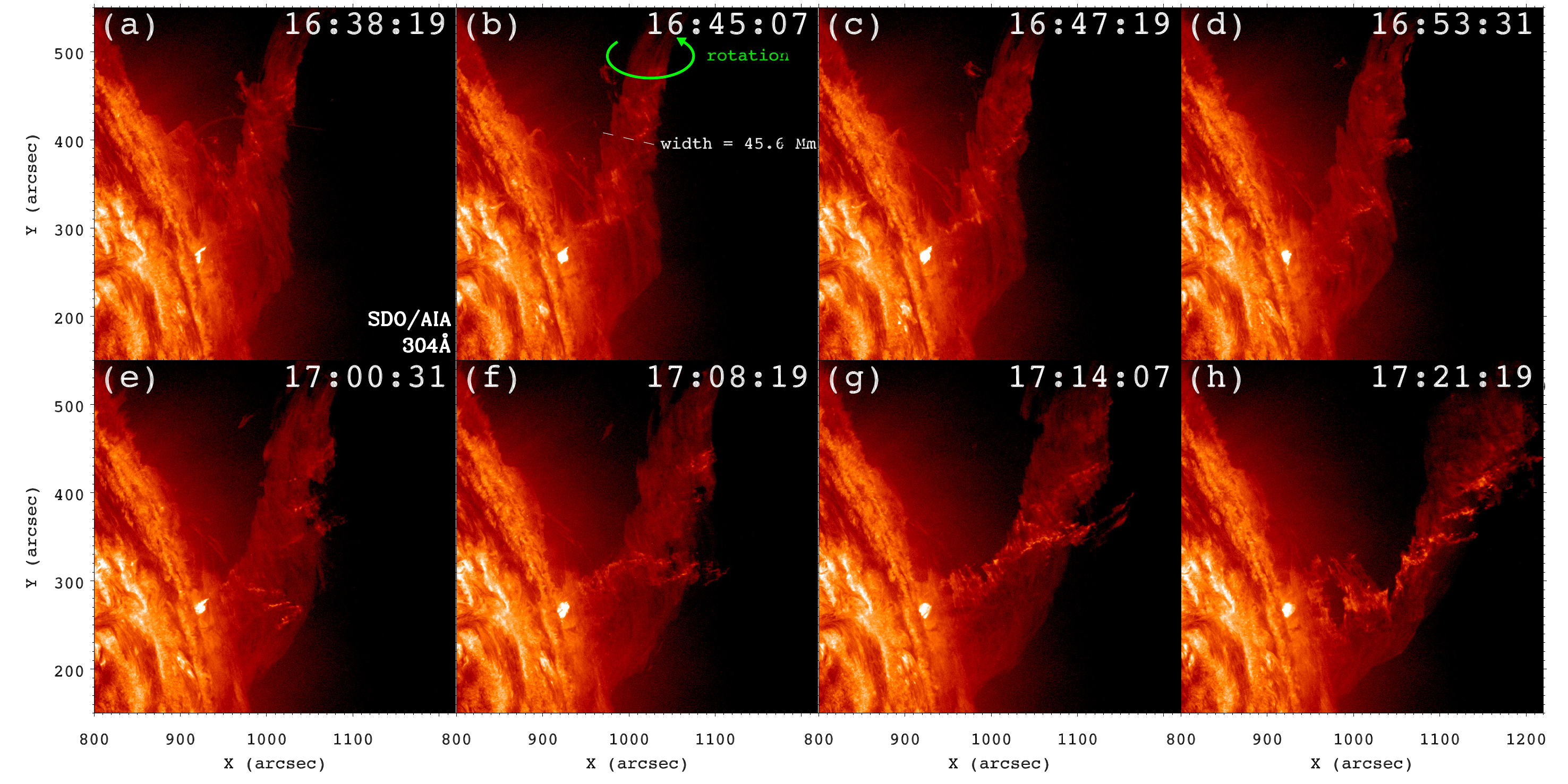}}
\caption{Snapshots of the rotating prominence observed by AIA in 304 {\AA} during 16:38$-$17:21 UT. 
In panel (b),} the green arrow indicates the counterclockwise direction of rotation viewed from above.
The white dashed line is used to measure the width (45.6 Mm) of prominence.
\label{fig3}
\end{figure}

In Figure~\ref{fig2}, the top and bottom panels show 171 {\AA} and 304 {\AA} images observed by SDO/AIA (see also online animation \textsf{anim1.mp4}).
The prominence rises from the farside and appears above the western limb. Rapid rotation of the prominence body is obvious during prominence eruption (16:34$-$17:21 UT).
Figure~\ref{fig3} shows eight snapshots of the prominence during 16:38$-$17:21 UT.
In panel (b), the white dashed line across the prominence is used to measure the width ($D$) of prominence, being $\sim$45.6 Mm.
Assuming a cylindrical shape of the prominence, the radius is $\sim$22.8 Mm.

\begin{figure}
\centerline{\includegraphics[width=0.93\textwidth,clip=]{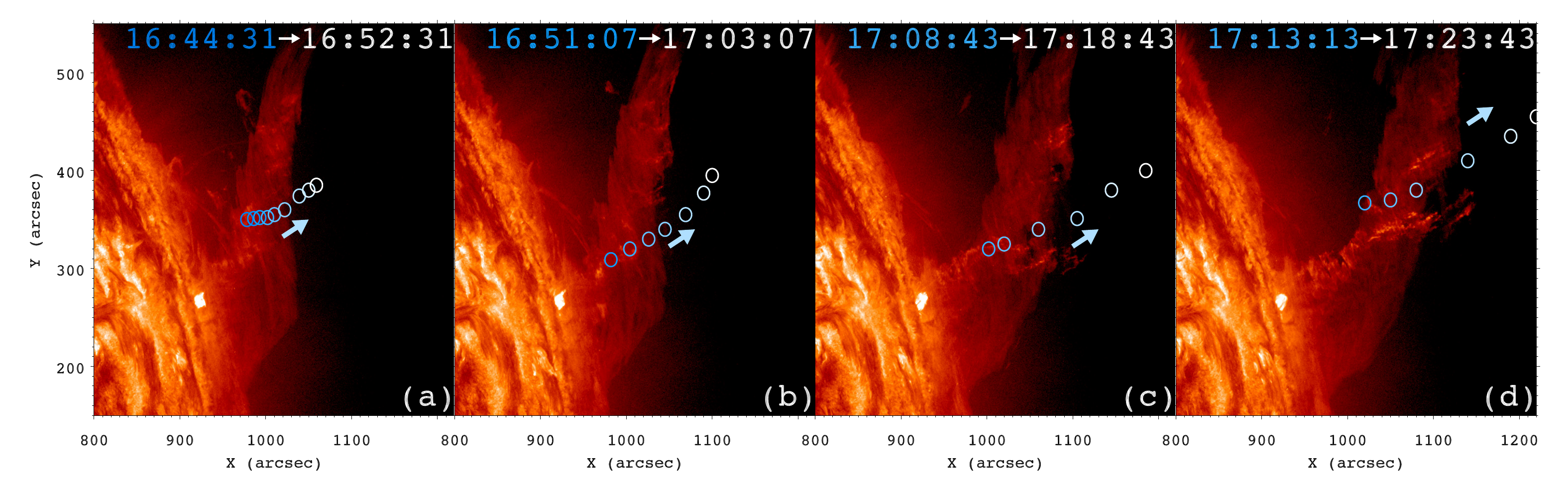}}
\caption{Snapshots of the prominence observed by AIA in 304 {\AA}. The circles denote the trajectories of bright features in the foreground during the rotation.
On top of each panel, the time in blue corresponds to the observing time of the EUV image.}
\label{fig4}
\end{figure}

\begin{figure}
\centerline{\includegraphics[width=0.8\textwidth,clip=]{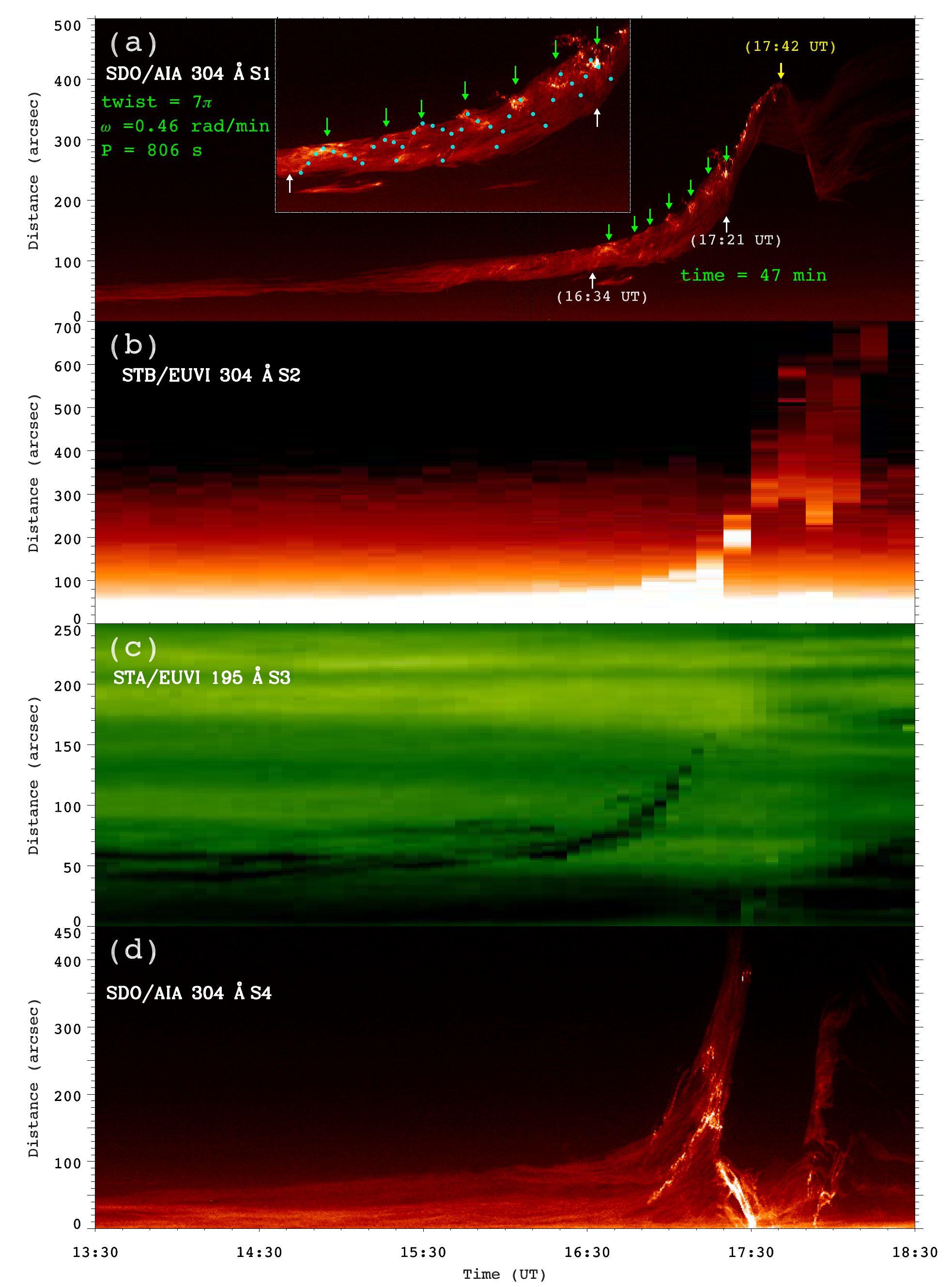}}
\caption{(a) Time-slice diagram of S1 in AIA 304 {\AA}. $s=0$ and $s=500\arcsec$ represent the northeast and southwest endpoints of S1, respectively.
The white arrows indicate the start and end times of prominence rotation, which lasts for $\sim$47 minutes. The green arrows point to the bright features. 
The yellow arrow denotes the turning point when the direction of drift reverses. 
The inset shows an enlarged view of the time-slice diagram during rotation. The light blue dots outline the sinusoidal patterns.}
The values of total twist ($\Phi=7\pi$), average angular speed ($\omega=0.46$ rad min$^{-1}$) and period ($P=806$ s) are labeled.
(b) Time-slice diagram of S2 in STB/EUVI 304 {\AA}. (c) Time-slice diagram of S3 in STA/EUVI 195 {\AA}. (d) Time-slice diagram of S4 in AIA 304 {\AA}.
\label{fig5} 
\end{figure}

The online movie (\textsf{anim1.mp4}) indicates that the direction of rotation is counterclockwise viewed from above.
Figure~\ref{fig4} shows snapshots of the prominence observed by AIA in 304 {\AA}. 
We tracked some of the bright features in the foreground during rotation, which are marked with blue and white circles.
It is clear that the prominence is rotating counterclockwise.
Contrary to the rotating jets whose axes are almost static \citep[e.g.,][]{Cur2011,Shen2011b,Chen2012,Hong2013,Sch2013,Zha2014}, the prominence rises and drifts southwestward during rotation.
To investigate the rotation in detail, we choose a curved slice (S1) across the leg in Figure~\ref{fig2}(b1).
The time-slice diagram of S1 in 304 {\AA} is displayed in Figure~\ref{fig5}(a). The sinusoidal patterns (light blue dots) in the inset indicate rapid rotation during 16:34$-$17:21 UT \citep{Oka2016}.
Therefore, the rotation lasts for $\sim$47 minutes. The seven green arrows point to the bright features that appear periodically.
The corresponding 304 {\AA} images at these moments are displayed in Figure~\ref{fig3}.
The average period ($P$) is calculated to be $\sim$806 s, and the corresponding angular speed ($\omega$) is $\sim$0.46 rad min$^{-1}$.
Considering the width of the prominence in Figure~\ref{fig3}(b), the linear speed is $v=\omega D/2\approx355$ km s$^{-1}$.
The total angle of rotation reaches $\sim$7$\pi$, which significantly exceeds the critical twist for kink instability \citep{Hood1981,Tor2003,Tor2004}.
Accordingly, the rapid rotation is indicative of untwisting motion of a highly twisted prominence.
In Figure~\ref{fig5}(a), the yellow arrow denotes the turning point ($\sim$17:42 UT) when the direction of drift reverses, which implies the writhing motion \citep{Fan2003,Tor2005,Liu2007}.
It should be noted that rotation in the clockwise direction near the footpoint of prominence is observed during 17:40$-$18:05 UT. 
Such reversal of direction of prominence rotation has been reported by \citet{Thom2012} and \citet{Song2018}.
This short period of rotation is not considered when calculating the total twist of the prominence.

To investigate the rising motion of the prominence, we choose a straight slice (S4) in Figure~\ref{fig2}(b3). The time-slice diagram of S4 in 304 {\AA} is displayed in Figure~\ref{fig5}(d).
The prominence undergoes a prolonged slow rise phase and a fast rise phase. The fast rise starts from $\sim$16:50 UT until 17:30 UT when the upper part escapes the FOV of AIA to drive a CME.

\begin{figure}
\centerline{\includegraphics[width=0.9\textwidth,clip=]{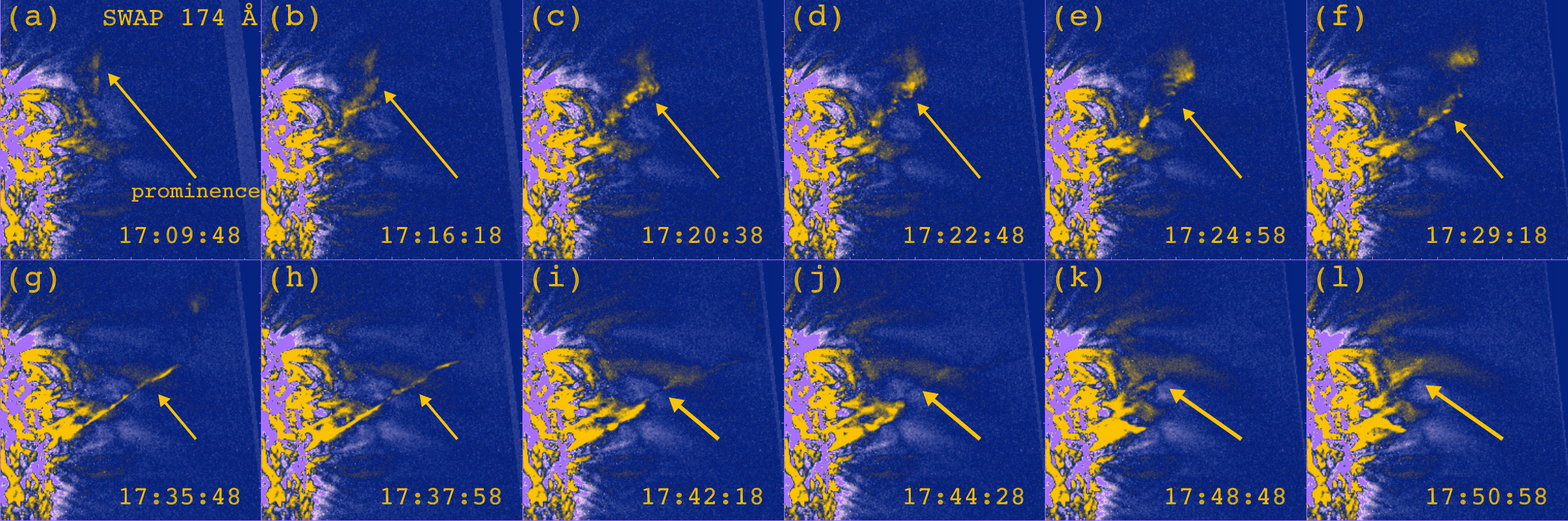}}
\caption{Base-difference images of the erupting prominence observed by SWAP in 174 {\AA}. The FOV is 800$\arcsec\times800\arcsec$.
The yellow arrows point to the prominence. An animation of this figure is available in the Electronic Supplementary Material (\textsf{anim2.mp4}).}
\label{fig6}
\end{figure}

The eruption was captured by SWAP in 174 {\AA} with a larger FOV. Figure~\ref{fig6} shows a series of base-difference images (see also the online animation \textsf{anim2.mp4}).
The prominence, which is indicated by the arrows, starts to rise at $\sim$17:09 UT, accompanied by southwestward drift until $\sim$17:42 UT (panels (a-i)).
Afterward, the direction of drift reverses, suggesting the writhing motion (panels (j-l)).

\begin{figure}
\centerline{\includegraphics[width=0.9\textwidth,clip=]{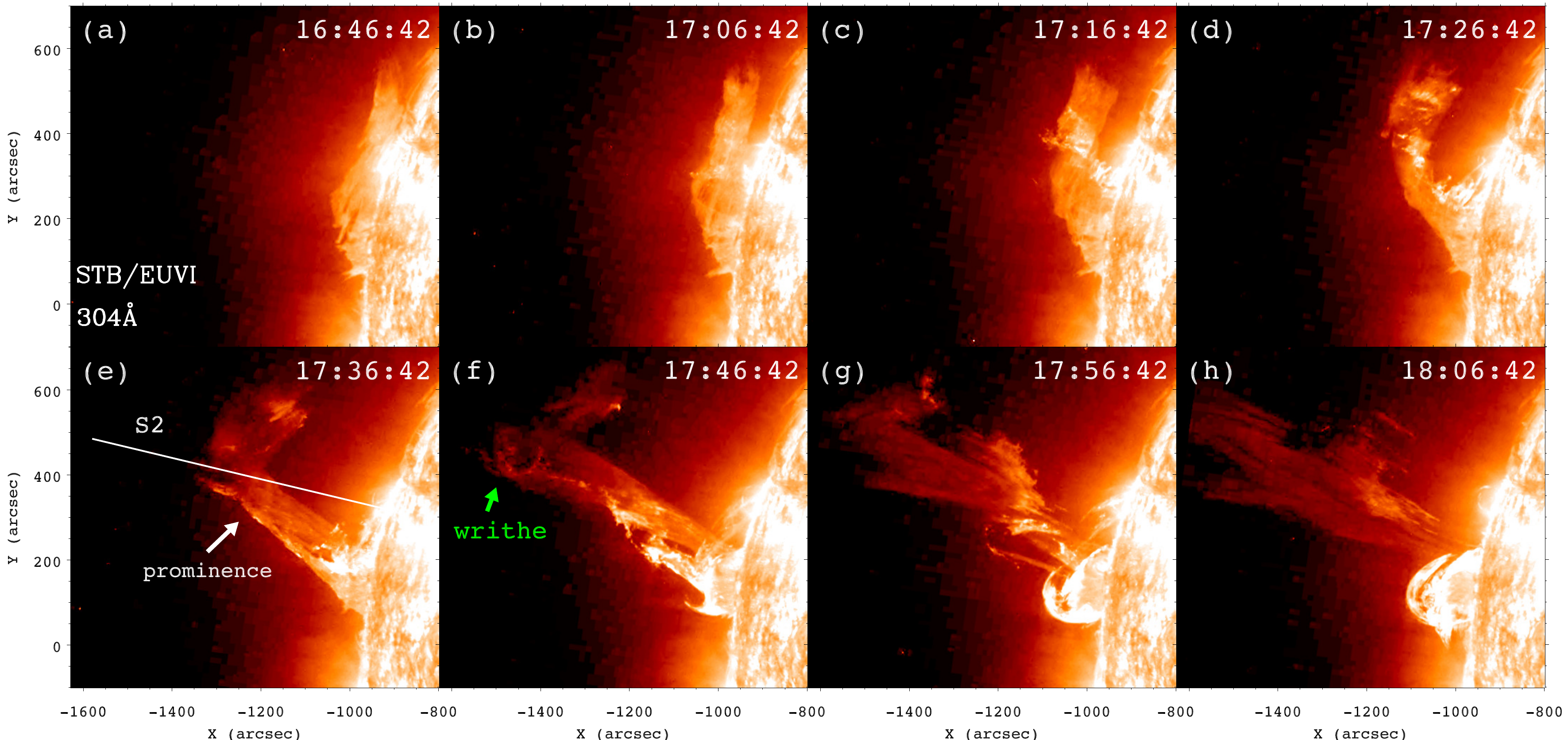}}
\caption{EUV 304 {\AA} images of the erupting prominence observed by STB/EUVI.
In panel (e), a straight slice (S2) is used to investigate the height evolution of prominence. In panel (f), the green arrow indicates the writhe of prominence.
An animation of this figure is available in the Electronic Supplementary Material (\textsf{anim3.mp4}).}
\label{fig7}
\end{figure}

As mentioned in Sect.~\ref{s-Intro}, the twin STEREO spacecraft observed the event from two perspectives.
Figure~\ref{fig7} shows a series of 304 {\AA} images observed by STB/EUVI (see also the online animation \textsf{anim3.mp4}). The prominence, which is indicated by the white arrow, 
elevates slowly from $\sim$17:00 UT and experiences a clear writhing motion around 17:46 UT with an ``inverted $\gamma$" shape, which is indicated by the green arrow in panel (f).
The top of the prominence when kink instability takes place at 17:46:42 UT is $\sim$405\,Mm (0.58\,$R_\odot$) above the solar limb. 
Using triangulation method, the true height is estimated to be $\sim$488\,Mm (0.70\,$R_\odot$).
Afterward, the prominence continues to rise and escapes the FOV of STB/EUVI (panels (g-h)).
To investigate the height evolution, a straight slice (S2) is selected in panel (e). The time-slice diagram of S2 is plotted in Figure~\ref{fig5}(b).
The height evolution is also characterized by a slow rise phase and a fast rise phase, which are in agreement with the trend in AIA 304 {\AA} (Figure~\ref{fig5}(d)).

\begin{figure}
\centerline{\includegraphics[width=0.9\textwidth,clip=]{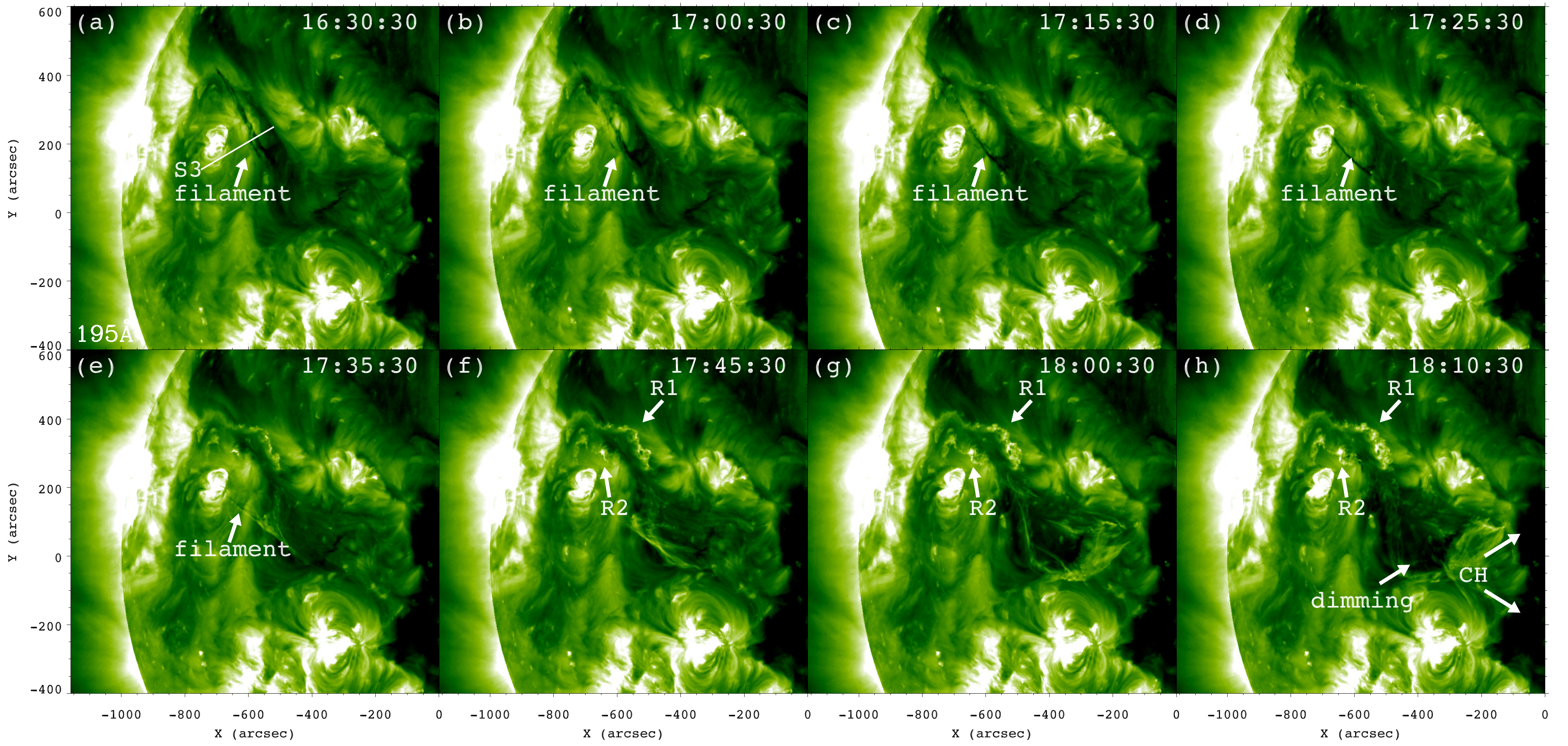}}
\caption{EUV 195 {\AA} images of the rising filament observed by STA/EUVI.
The arrows point to the dark filament, bright flare ribbons (R1 and R2), hooklike dimming, and an equatorial coronal hole (CH).}
\label{fig8}
\end{figure}

The bright prominence manifests itself as a dark filament in the FOV of STA/EUVI. Figure~\ref{fig8} shows 195 {\AA} images observed by STA/EUVI.
The filament starts to rise at $\sim$17:00 UT until $\sim$17:35 UT when the spine of filament becomes undistinguishable (panel (e)).
The southwest footpoint of the filament drifts and generates a hooklike dimming region close to an equatorial coronal hole \citep{Wang2017,Aul2019,Lor2021}.
The eruption generates a pair of bright flare ribbons (R1 and R2), which undergo separation.

\begin{figure}
\centerline{\includegraphics[width=0.9\textwidth,clip=]{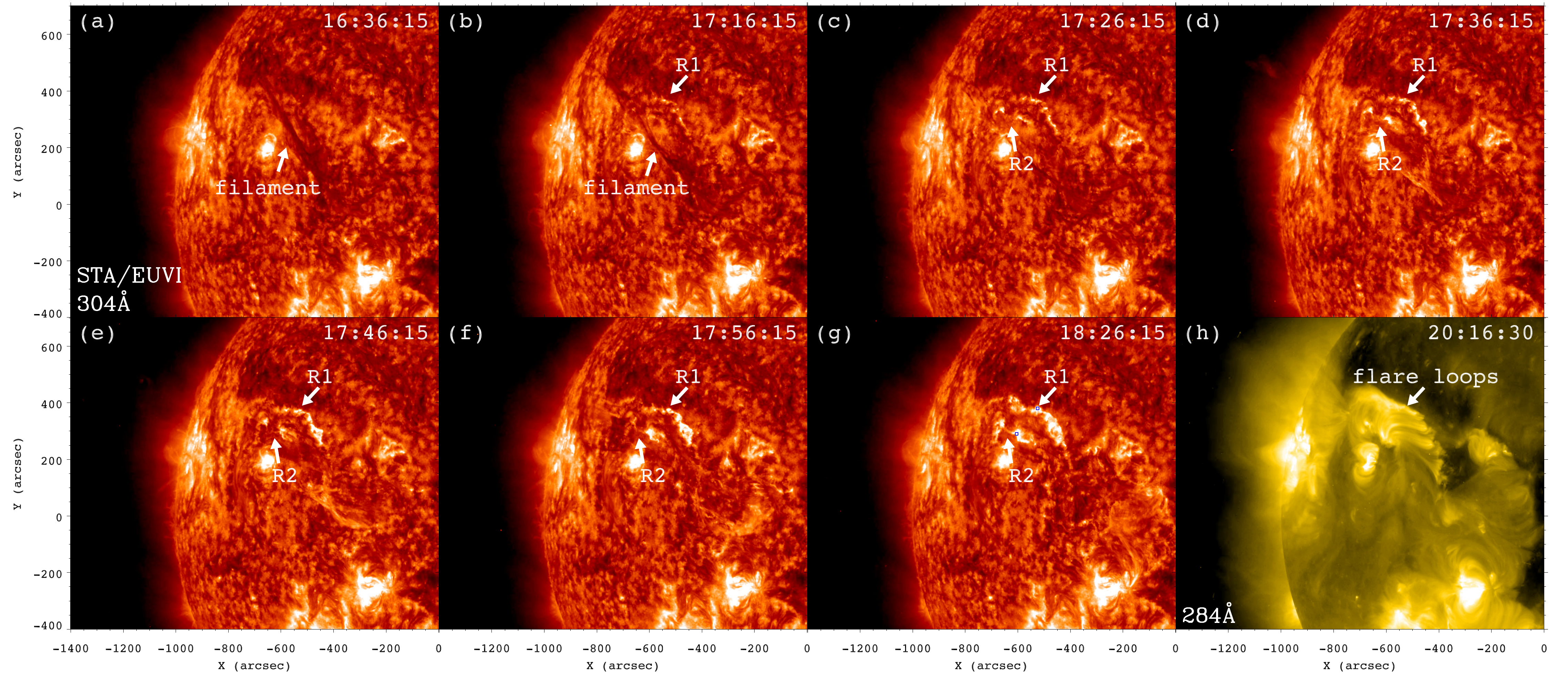}}
\caption{EUV 304 and 284 {\AA} images of the rising filament observed by STA/EUVI.
The arrows point to the dark filament, bright flare ribbons (R1 and R2), and post flare loops.
In panel (g), two blue boxes are used to calculate the total intensities of R1 and R2.}
\label{fig9}
\end{figure}

\begin{figure}
\centerline{\includegraphics[width=0.6\textwidth,clip=]{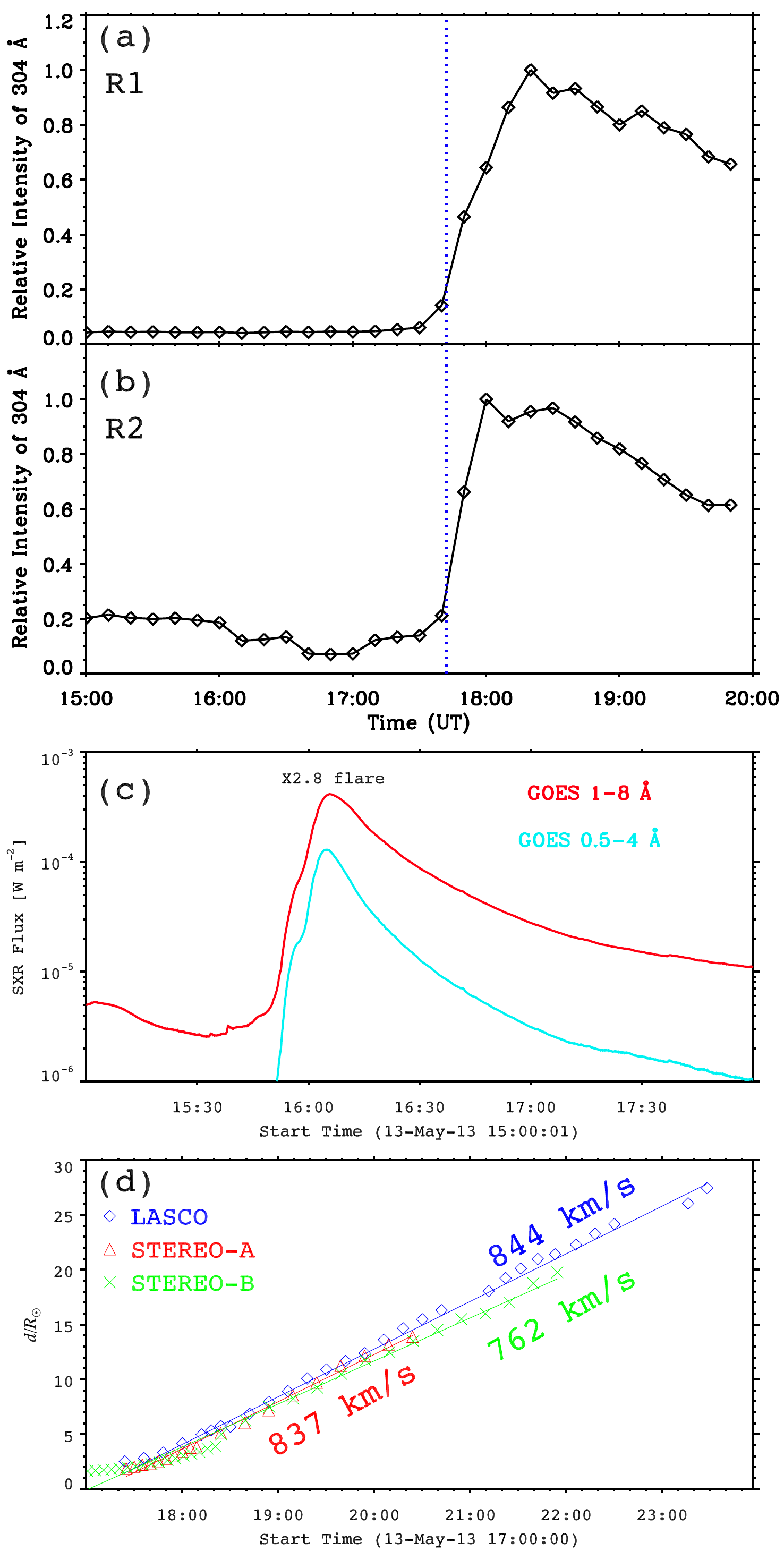}}
\caption{(a-b) Normalized total intensities of the flare ribbons (R1 and R2). The blue dashed line denotes the onset time ($\sim$17:42 UT) of kink instability.
(c) SXR light curves during 15:00$-$18:00 UT.
(d) Height evolutions of the CME LE observed by LASCO (blue diamonds), STA (red triangles), and STB (green crosses), respectively.
The corresponding linear speeds are labeled.}
\label{fig10}
\end{figure}

Figure~\ref{fig9} shows 304 and 284 {\AA} images observed by STA/EUVI. The evolution of filament and flare ribbons in 304 {\AA} is analogous to that in 195 {\AA}.
The bright post flare loops connecting the ribbons are captured in 284 {\AA} ($\log T\approx6.3$) in panel (h).
In panel (g), two representative boxes ($3\arcsec\times3\arcsec$) at the conjugate flare ribbons are selected. The normalized total intensities of the boxes are plotted in Figure~\ref{fig10}(a-b).
The EUV intensities increase from $\sim$17:30 UT and peak at 18:00$-$18:20 UT followed by a gradual decay.
The blue dashed line denotes the onset time ($\sim$17:42 UT) of kink instability, which is consistent with the beginning of the flare impulsive phase.
For comparison, the SXR light curves in 1$-$8 {\AA} and 0.5$-$4 {\AA} during 15:00$-$18:00 UT are plotted with red and cyan lines in Figure~\ref{fig10}(c).
The huge enhancements of SXR emission starting at 15:48 UT and peaking at 16:05 UT originate from the X2.8 class flare in NOAA active region 11748 close to the eastern limb \citep{Gou2020}.
There is no SXR response of the flare ribbons observed in 304 {\AA} by STA/EUVI, meaning that the flare completely occurred at the farside.
 
\begin{figure}
\centerline{\includegraphics[width=0.8\textwidth,clip=]{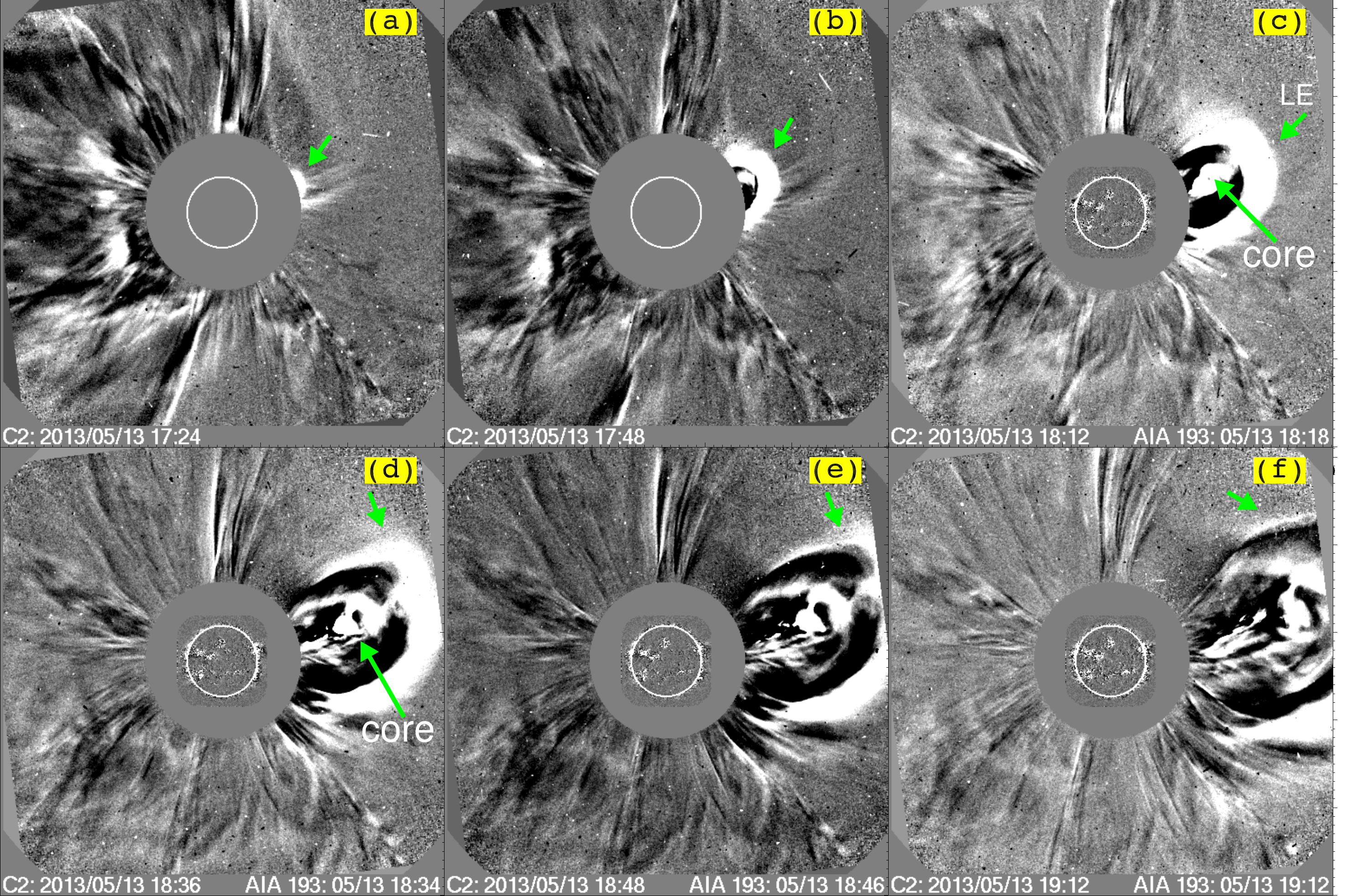}}
\caption{Running-difference images of the CME observed by LASCO-C2. The green arrows indicate the CME leading edge (LE) and core.}
\label{fig11}
\end{figure}

\begin{figure}
\centerline{\includegraphics[width=0.8\textwidth,clip=]{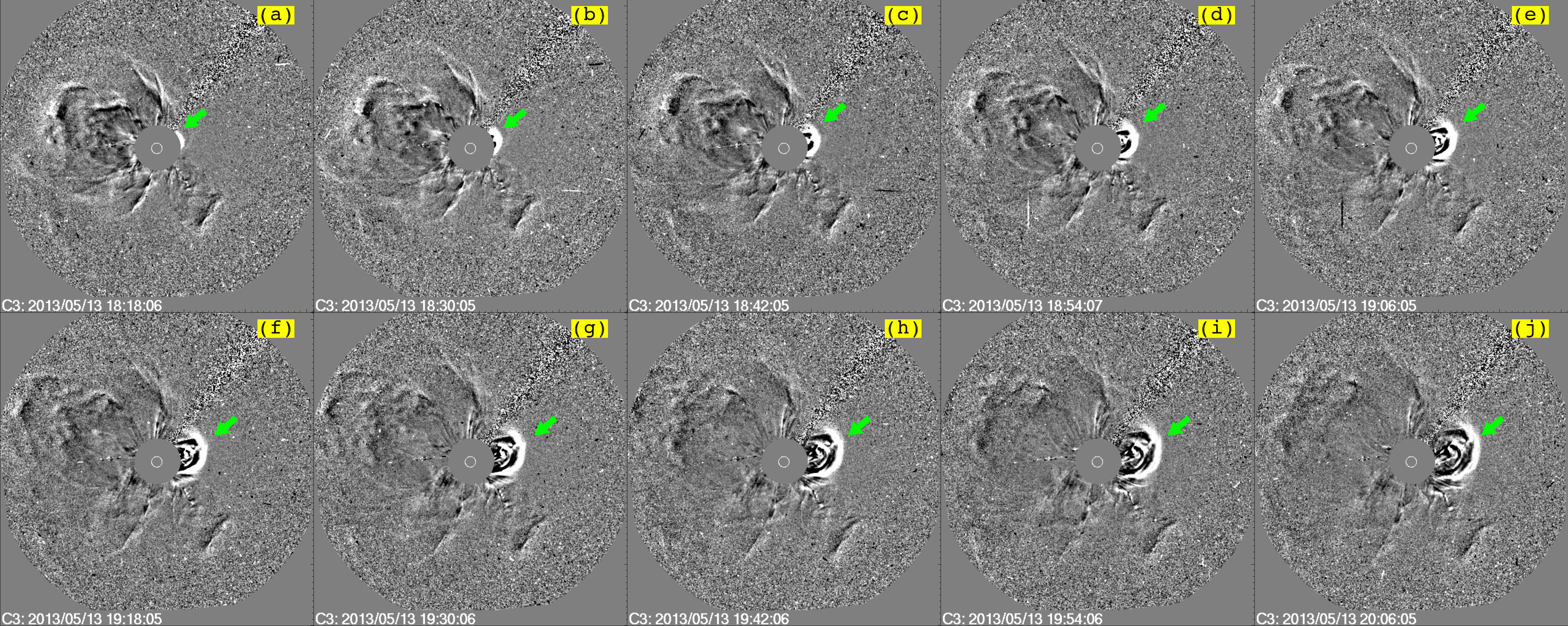}}
\caption{Running-difference images of the CME observed by LASCO-C3. The green arrows indicate the CME leading edge (LE).}
\label{fig12}
\end{figure}

As mentioned in Sect.~\ref{s-Intro}, the eruption drives a fast CME. The running-difference images of the CME observed by LASCO-C2 are displayed in Figure~\ref{fig11}.
The CME first appears at $t_{0}=$17:24 UT and rises up in the northwest direction, showing a typical three-part structure: a bright leading edge, a dark cavity, and a bright core (panels (c-d)).
The core is widely believed to consist of an erupting prominence, which is presented in Figure~\ref{fig2} \citep{Liu2007,Song2022}.
The CME propagates further until 20:06 UT in the FOV of LASCO-C3 (Figure~\ref{fig12}), and the height evolution is drawn with blue diamonds in Figure~\ref{fig10}(d). 
The fitted linear speed ($V$) is $\sim$844 km s$^{-1}$.
The parameters, including $t_0$, position angle (PA), final angular width (AW), and $V$ of the CME observed by LASCO, are listed in the second column of Table~\ref{tab-2}.

\begin{table}
\caption{Parameters of the CME observed by the WL coronagraphs on board LASCO, STB, and STA. 
For comparison, the values of AW and $V$ from the GCS modeling are listed in the last column.}
\label{tab-2}
\tabcolsep 1.5mm
\begin{tabular}{ccccc}
  \hline
Spacecraft & LASCO   &  STB  &  STA & GCS \\ 
Instrument &  C2, C3   &  COR1, COR2    &  COR1, COR2 & $-$  \\
  \hline
$t_0$ (UT) & 17:24 & 17:10 &  17:25 & $-$ \\
PA ($^{\circ}$) & $\sim$288 & $\sim$72 & $\sim$74 & $-$ \\
AW ($^{\circ}$) & $\sim$82 & $\sim$77 & $\sim$82 & $\sim$114 \\
$V$ (km s$^{-1}$) & $\sim$844 & $\sim$762 & $\sim$837 & $\sim$1005 \\
  \hline
\end{tabular}
\end{table}

\begin{figure}
\centerline{\includegraphics[width=0.8\textwidth,clip=]{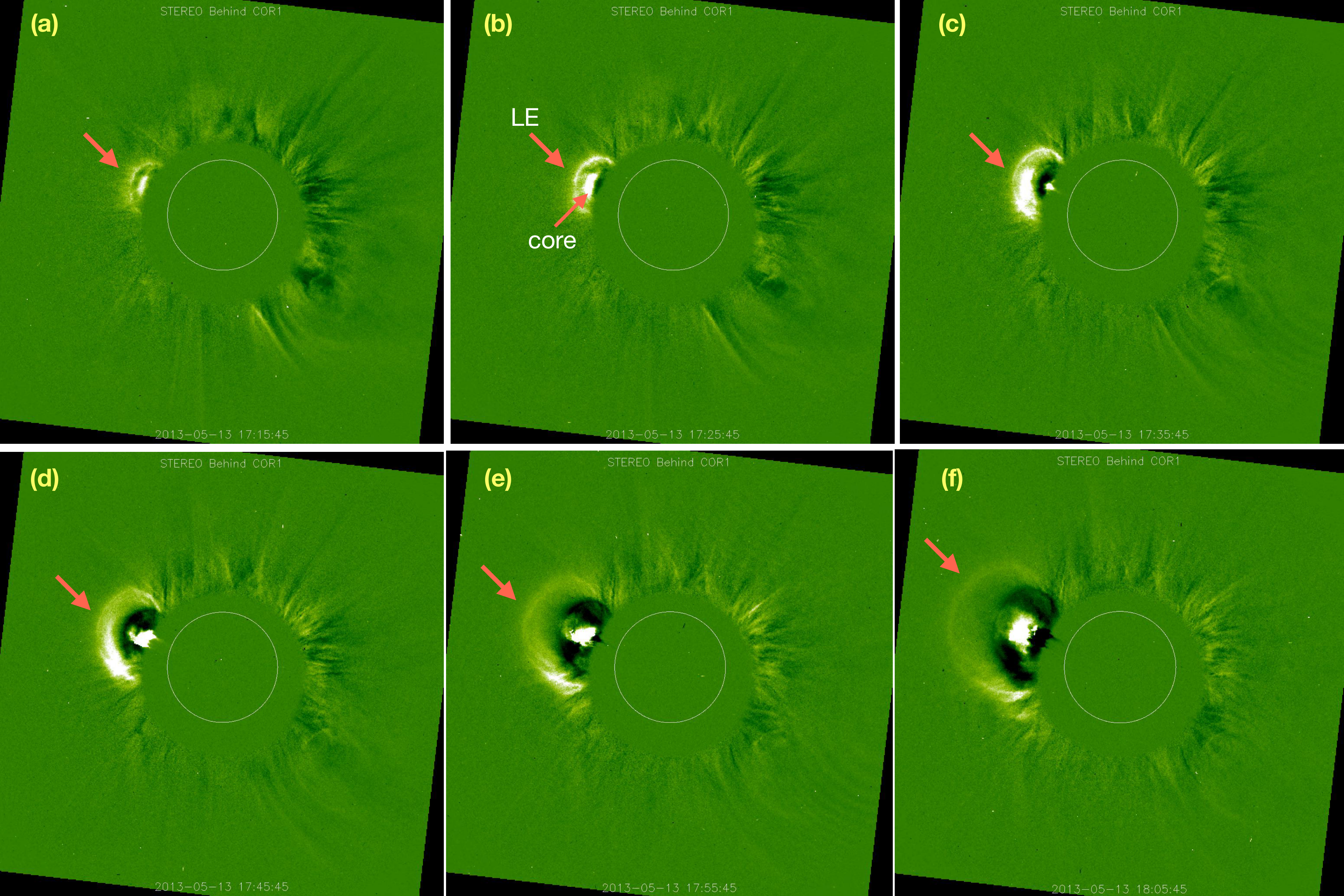}}
\caption{Running-difference images of the CME observed by STB/COR1. The orange arrows indicate the CME leading edge (LE) and core.}
\label{fig13}
\end{figure}

\begin{figure}
\centerline{\includegraphics[width=0.8\textwidth,clip=]{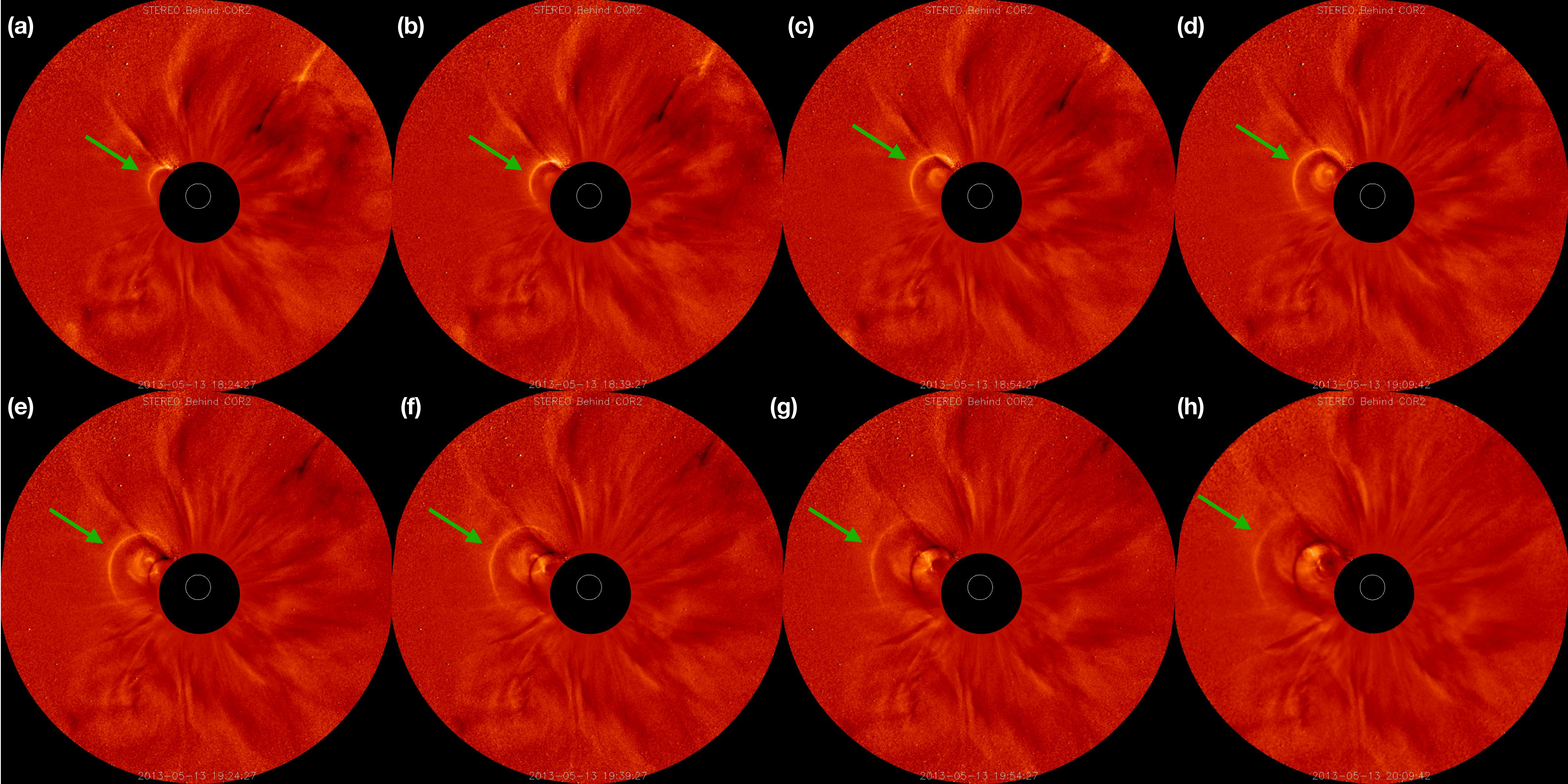}}
\caption{Running-difference images of the CME observed by STB/COR2. The green arrows indicate the CME leading edge (LE).}
\label{fig14}
\end{figure}

\begin{figure}
\centerline{\includegraphics[width=0.8\textwidth,clip=]{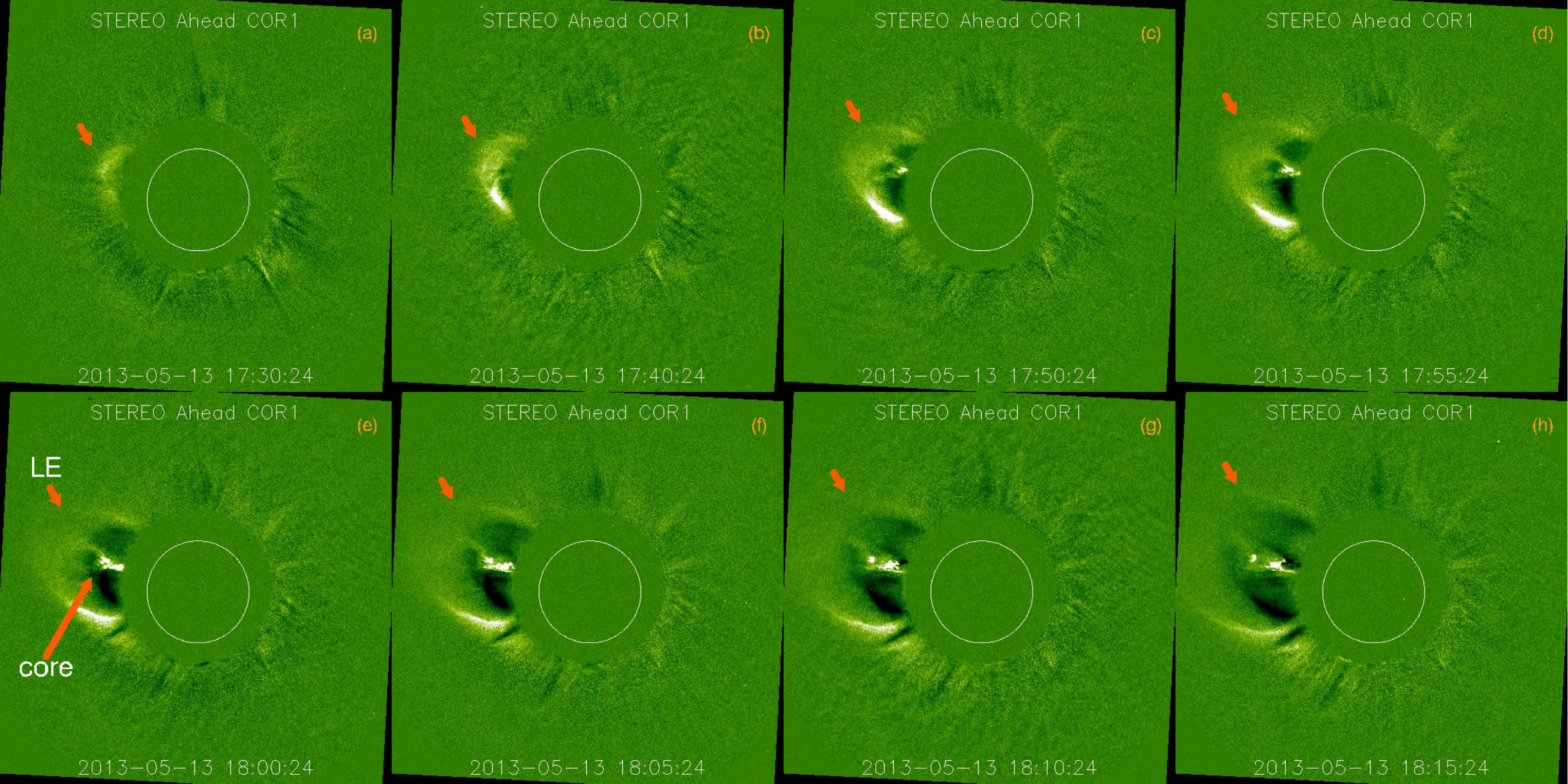}}
\caption{Running-difference images of the CME observed by STA/COR1. The orange arrows indicate the CME leading edge (LE) and core.}
\label{fig15}
\end{figure}

\begin{figure}
\centerline{\includegraphics[width=0.8\textwidth,clip=]{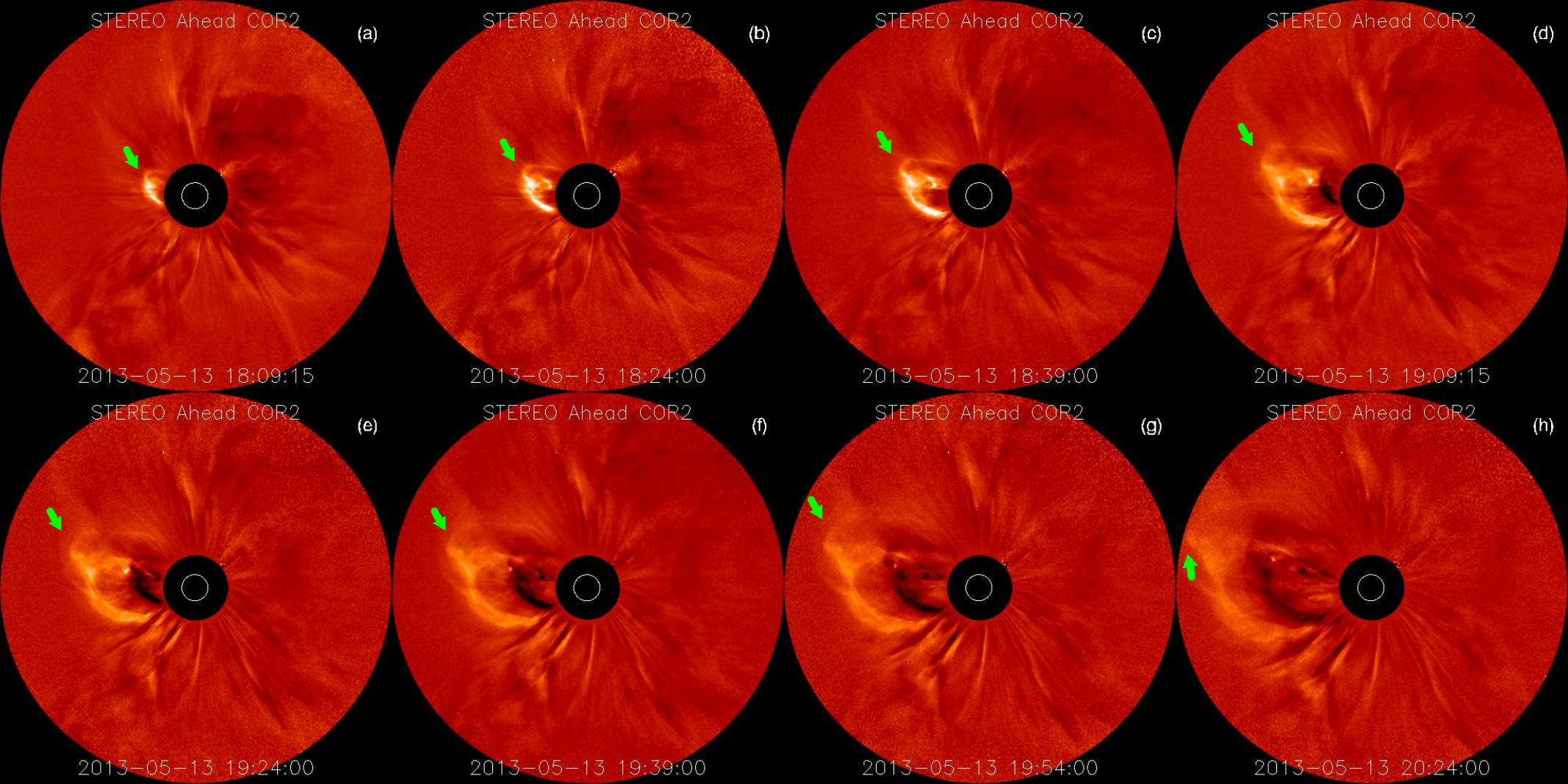}}
\caption{Running-difference images of the CME observed by STA/COR2. The green arrows indicate the CME leading edge (LE).}
\label{fig16}
\end{figure}

Likewise, the running-difference images of the CME in the FOV of COR1 and COR2 on board STB are displayed in Figure~\ref{fig13} and Figure~\ref{fig14}, respectively.
The CME first appears at 17:10 UT with the same three-part morphology and propagates in the northeast direction until 20:09 UT at a speed of $\sim$762 km s$^{-1}$ 
(see the third column of Table~\ref{tab-2}).
The running-difference images of the CME in the FOV of COR1 and COR2 on board STA are displayed in Figure~\ref{fig15} and Figure~\ref{fig16}, respectively.
The CME first appears at 17:25 UT and propagates in the northeast direction until 20:24 UT at a speed of $\sim$837 km s$^{-1}$ (see the fourth column of Table~\ref{tab-2}).
The WL intensities of CME get too weak to be identified after 20:30 UT. After checking the WL images of the CME carefully, we conclude that the CME is not a Cartwheel CME, 
which carries away a part of the twist and consequently shows highly visible rotation around the line of sight \citep{Thom2012,Pant2018}.

\subsection{3D Reconstruction of the CME using GCS modeling} \label{s-gcs}
Simultaneous observations of the CME enable us to perform a 3D reconstruction of the CME. Although the CME LE shows a nearly symmetric bubble in the FOVs of LASCO and STB, 
it looks raised and irregular in the FOV of STA/COR2, which is unsuitable for cone models \citep{Mich2003,Zha2021}.
Considering the helical structure and rapid rotation of the prominence (Figures~\ref{fig2} and \ref{fig3}), it is concluded that the CME is driven by a highly twisted magnetic flux rope.
Naturally, we turn to the GCS modeling \citep{The2006}. The GCS, resembling a croissant, is composed of two conical ``legs" with angular separation of 2$\alpha$ 
and a circular annulus of varying radius:
\begin{equation} \label{eqn-3}
 a(r)=\kappa r,
\end{equation}
where $r$ is the distance from the Sun center to a point at the outer edge of the shell, and $\kappa$ is the CME aspect ratio.
Each leg has a height of $h$ and half angle of $\delta$ \citep{The2011}:
\begin{equation} \label{eqn-4}
 \kappa=\sin\delta.
\end{equation}
Two auxiliary parameters are:
\begin{equation} \label{eqn-5}
 b=h/\cos\alpha, 
\end{equation}
\begin{equation} \label{eqn-6}
 \rho=h\tan\alpha.
\end{equation}
Accordingly, the heliocentric height of the LE is:
\begin{equation} \label{eqn-7}
 h_{\mathrm{LE}}=\frac{b+\rho}{1-\kappa}=\frac{h}{1-\kappa}(\frac{1}{\cos\alpha}+\tan\alpha).
\end{equation}
The edge-on and face-on angular widths are:
\begin{equation} \label{eqn-8}
 \omega_{\mathrm{EO}}=2\delta,
\end{equation}
\begin{equation} \label{eqn-9}
 \omega_{\mathrm{FO}}=2(\alpha+\delta)=2\alpha+\omega_{\mathrm{EO}}.                    
\end{equation}
The source region of the CME has a Carrington longitude $\phi$ and a latitude $\theta$. The tilt angle of the PIL is denoted by $\gamma$.
Since the eruption of large-scale prominence occurs at the farside, the value of $\phi$ is changed repeatedly and eventually set to be 91$^{\circ}$.
The value of $\theta$ changes slightly to derive a better fitting, although it should be fixed if the CME does not experience a deflection (see Figure~\ref{fig19}(c)).
The value of $\gamma=45^{\circ}$ is roughly determined from the STA/EUVI images in Figure~\ref{fig9}. 
The value of $\kappa$ is set to be 0.5 ($\delta=30^{\circ}$) after many trials.
As the CME expands laterally, the value of $\alpha$ increases.

\begin{figure}
\centerline{\includegraphics[width=0.9\textwidth,clip=]{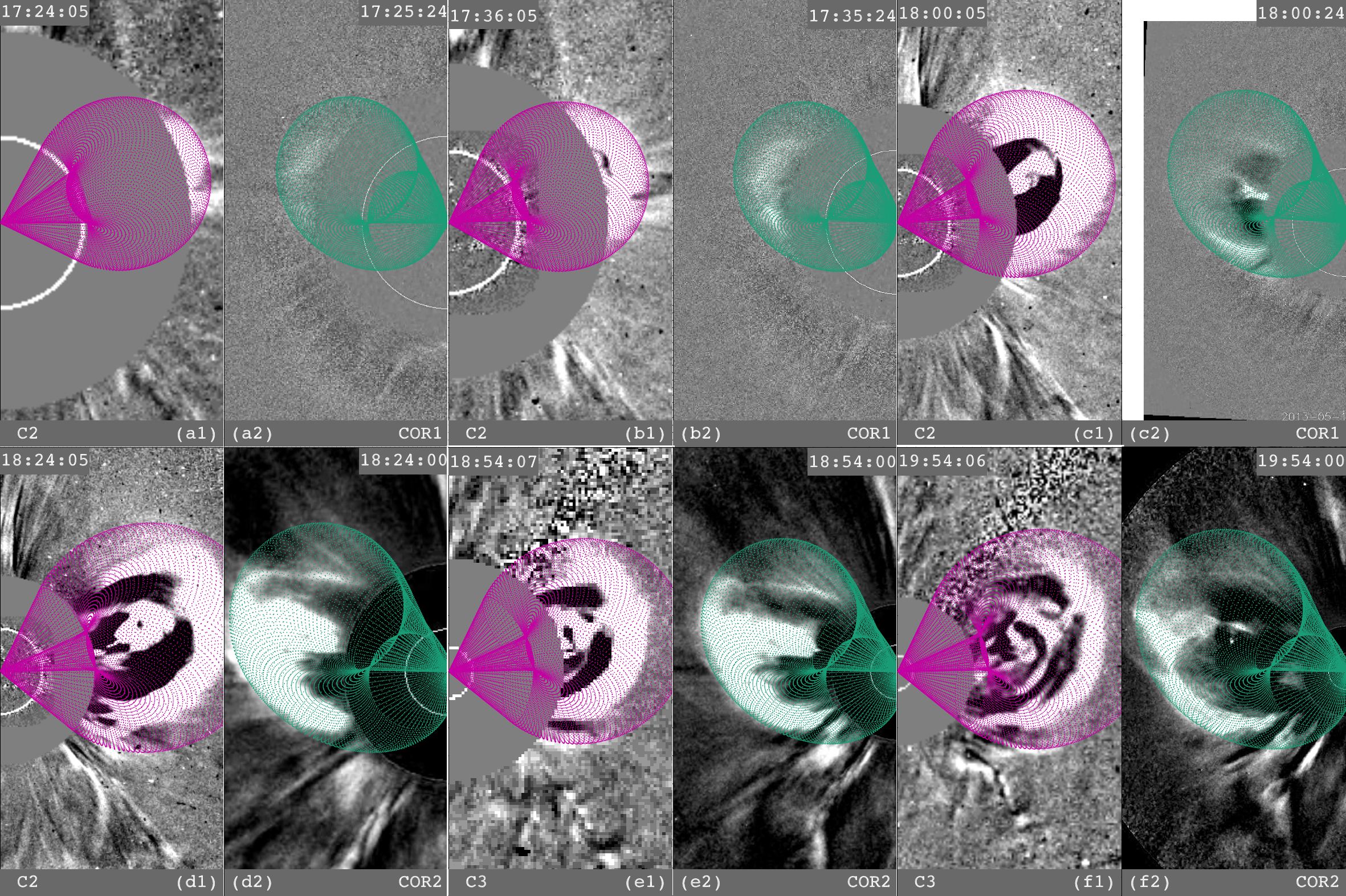}}
\caption{Selected WL images of the CME observed by LASCO and STA almost simultaneously, which are overlaid with reconstructed GCS models (red-violet and blue-green dots).}
\label{fig17}
\end{figure}

\begin{figure}
\centerline{\includegraphics[width=0.9\textwidth,clip=]{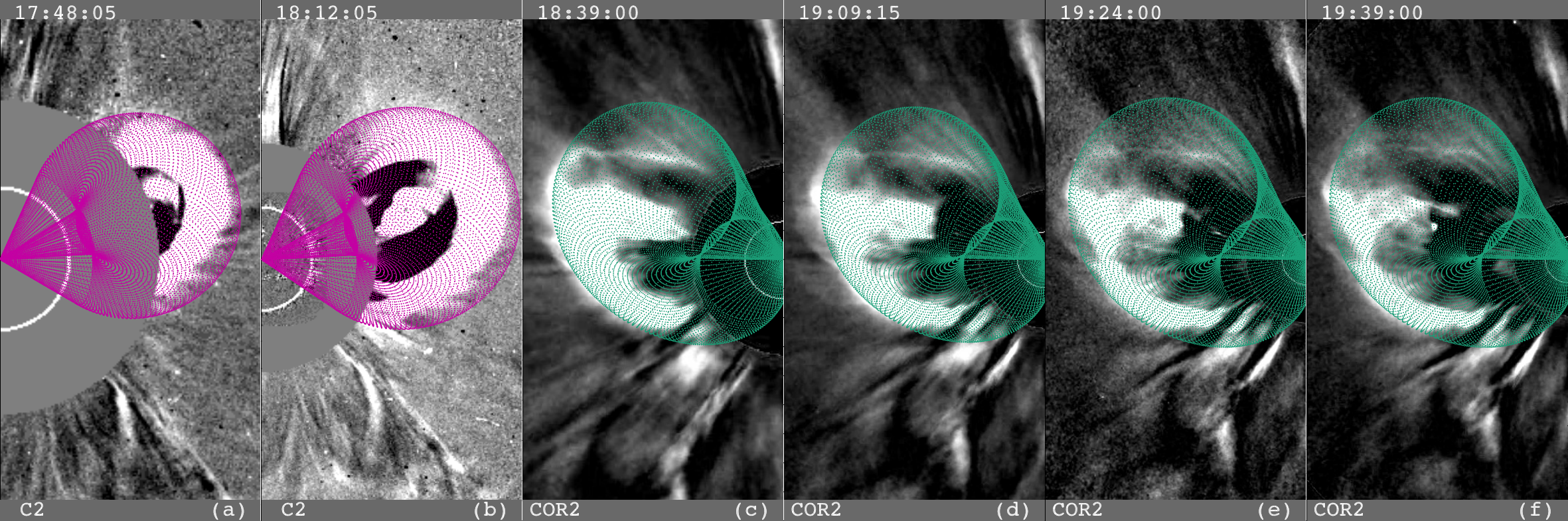}}
\caption{Selected WL images of the CME observed by LASCO (a-b) and STA (c-f), which are overlaid with reconstructed GCS models (red-violet and blue-green dots).}
\label{fig18}
\end{figure}

\begin{figure}
\centerline{\includegraphics[width=0.9\textwidth,clip=]{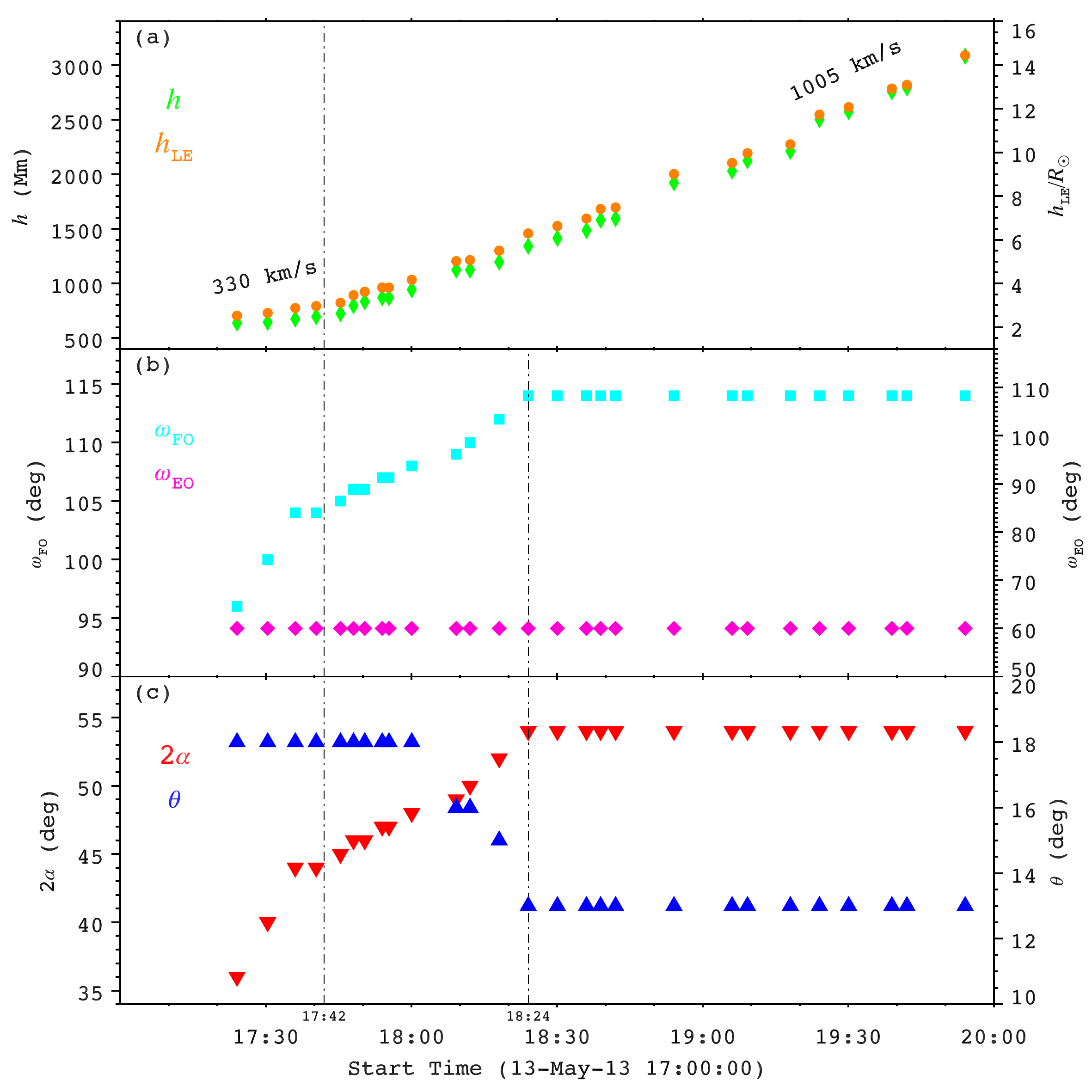}}
\caption{Time evolutions of the leg height ($h$), leading ledge height ($h_{\mathrm{LE}}$), 
face-on angular width ($\omega_{\mathrm{FO}}$), edge-on angular width ($\omega_{\mathrm{EO}}$), 
angular separation (2$\alpha$) of the legs, and latitude ($\theta$).
The left dot-dashed line denotes the onset time (17:42 UT) of the writhing motion. 
The right dot-dashed line denotes the time (18:24 UT) when 2$\alpha$ and $\omega_{\mathrm{FO}}$ reach the maxima.
In panel (a), the linear speeds before and after 17:42 UT are labeled.}
\label{fig19}
\end{figure}

To perform GCS modeling, we use WL images of the CME observed by LASCO and STA. 
Considering the inconsistency in time cadence between the two spacecrafts, we first use images, preferably at the same time.
Figure~\ref{fig17} shows a series of images of the CME observed by LASCO and STA during 17:24$-$19:54 UT. 
The reconstructed GCS models are projected with red-violet and blue-green dots, respectively. 
It is clear that the GCS models fit well with the observations. The leading edges of the CME are perfectly consistent with those of GCS models.
Time evolutions of the height ($h$), face-on angular width ($\omega_{\mathrm{FO}}$), edge-on angular width ($\omega_{\mathrm{EO}}$), 2$\alpha$, and $\theta$ are plotted in Figure~\ref{fig19}.
For those not simultaneously observed by LASCO and STA, we use WL images from one perspective for reconstruction. 
Figure~\ref{fig18} shows a series of images of the CME observed by LASCO (a-b) and STA (c-f), which are overlaid with the GCS models (red-violet and blue-green dots), 
indicating that the fittings are still satisfactory.

The results of fittings are plotted in Figure~\ref{fig19}.
The separation angle of the legs (red triangles in panel (c)) and the face-on angular width (cyan squares in panel (b)) increase sharply from $\sim$17:24 UT to $\sim$17:36 UT, 
implying a quick lateral expansion \citep{Pats2010,Cre2020,Maj2020,Maj2022}.
Afterwards, the two parameters increase gradually until $\sim$18:24 UT and nearly keep constant (54$^{\circ}$ and 114$^{\circ}$). 
The maximum of $\omega_{\mathrm{FO}}$ is larger than the apparent widths of CME in each coronagraph (see Table~\ref{tab-2}).
The final ratio $\frac{\omega_{\mathrm{FO}}}{\omega_{\mathrm{EO}}}$ is equal to 1.9, which is consistent with the value reported by \citet{Kra2006}
and the average value (1.84$\pm$0.37) of the 12 events reported by \citet{Cre2020}.
Recently, \citet{Maj2022} investigated the rapid initial acceleration and width expansion of five CMEs at lower heights using the GCS modeling.
It is found that the face-on widths expand faster than the edge-on widths (see their Fig. 3), which is in agreement with our finding.
The latitude $\theta$ decreases slightly from 18$^{\circ}$ at 18:00 UT to 13$^{\circ}$ at 18:24 UT, 
suggesting an equatorward deflection during propagation \citep{Byr2010,Gui2011,Shen2011a,Liu2018,Maj2020}.

The leg height $h$ increases from $\sim$638 Mm at 17:24 UT to $\sim$3081 Mm at 19:54 UT, 
and the leading edge height $h_{\mathrm{LE}}$ increases from $\sim$2.5$R_{\odot}$ to $\sim$14.5$R_{\odot}$.
The final speed ($\sim$1005 km s$^{-1}$) in 3D is greater than the apparent speeds of CME in each coronagraph (see Figure~\ref{fig19}(a) and Table~\ref{tab-2}).
Although the CME is fast and wide enough, which is capable of driving an interplanetary shock \citep{Mor2019,Zha2022c}, 
no type II radio burst was detected by the ground-based radio stations\footnote{https://www.e-callisto.org}.
This is probably due to that the CME originates from the farside and propagates away from the Earth.

\begin{figure}
\centerline{\includegraphics[width=0.9\textwidth,clip=]{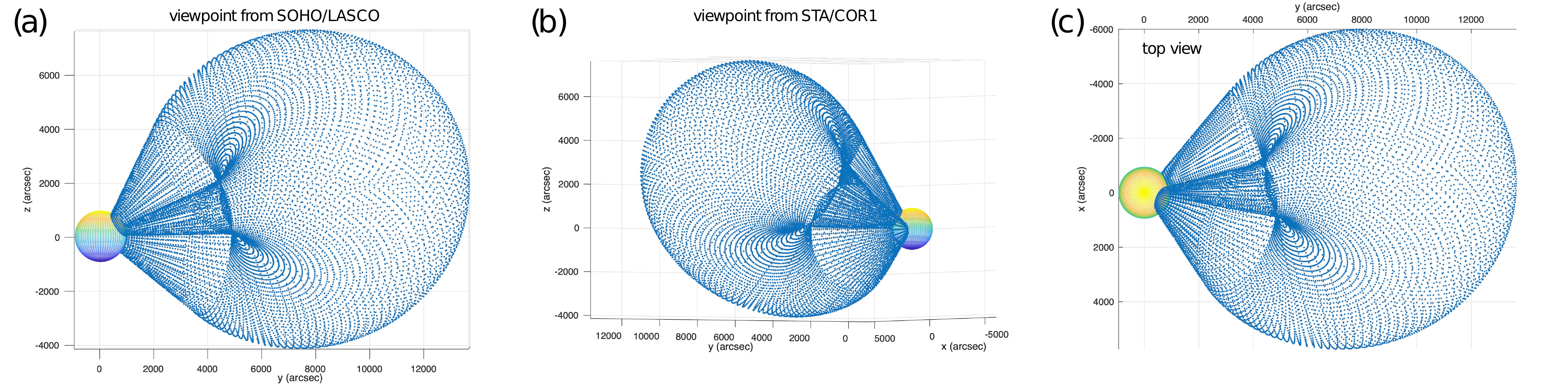}}
\caption{Viewpoints of the Sun (colored dots) and reconstructed GCS (blue dots) from SOHO/LASCO (a), STA/COR1 (b), and solar north at 19:54 UT, respectively.
An animation of this figure is available in the Electronic Supplementary Material (\textsf{anim4.mp4}).}
\label{fig20}
\end{figure}

In Figure~\ref{fig20}, the 3D view of the Sun (colored dots) and reconstructed GCS (blue dots) at 19:54 UT from perspectives of SOHO/LASCO (a), STA/COR1 (b), 
and solar north (c) are presented (see also the online animation \textsf{anim4.mp4}).

\section{Discussion} \label{s-Dis}
Prominences are rich in dynamics, such as large-amplitude oscillations \citep{Zha2012,Kuc2022}, upflows and downflows \citep{Ber2008}.
Rotations are also ubiquitous, not only in magnetized tornados \citep{Pet1943,Su2012,Su2014,Wed2012,Luna2015,Yang2018} 
but also in erupting prominences \citep{Thom2011,Thom2012,Liu2015,Song2018}. 
Interestingly, the direction of prominence rotation may reverse during the ascending phase \citep{Thom2012,Song2018}.
There are at least three reasons for rotation in erupting prominences. The first is ideal kink instability when the magnetic twist exceeds the threshold \citep{Yan2014b,Yan2014a}.
The second is twist transfer during magnetic reconnection between a highly twisted prominence and less twisted coronal loops \citep{Shi1986,Ku1987,Li2016,Oka2016,Xue2016,Yan2020}.
The third is the roll effect if a filament erupts non-radially \citep{Mar2003,Pan2008,Pan2013}. The top of the filament ribbon bends toward or away from the observer. 
As a consequence, the rolling motion propagates down the two legs of filaments simultaneously, generating significant twists in the legs with opposite chirality. 
That is to say, the two legs rotate in opposite directions (see Fig. 2 in \citet{Pan2013}).
Specifically, dextral filaments have a rolling motion toward the observer, which is accompanied by counterclockwise (clockwise) rotation at the left (right) leg, respectively.
Sinistral filaments have a rolling motion away from the observer, which is accompanied by counterclockwise (clockwise) rotation at the right (left) leg (see Fig. 4 in \citet{Pan2008}).
In Table~\ref{tab-3}, we compare the ideal kink instability with roll effect in detail, which are different at least in five aspects, including the shape of filament, initiation, 
chirality and rotation directions of two legs, and writhing motion.

\begin{table}
\caption{Comparison between the ideal kink instability and roll effect.}
\label{tab-3}
\tabcolsep 1.5mm
\begin{tabular}{ccc}
  \hline
 & kink instability  &  roll effect  \\ 
  \hline
shape of filament & twisted flux rope & ribbon-like \\
initiation &  pre-stored twist & bending at the top \\
chirality of two legs & same & opposite \\
rotation directions of two legs & same & opposite \\
writhing motion & yes & no \\
  \hline
\end{tabular}
\end{table}

In our study, the rising prominence keeps rotating in the counterclockwise direction for $\sim$47 minutes with an average period of $\sim$806 s (Figure~\ref{fig5}).
The total angle of rotation reaches 7$\pi$, which is significantly higher than the threshold of kink instability.
Besides, writhing motion is obviously observed by SWAP (Figure~\ref{fig6}) and STB/EUVI (Figure~\ref{fig7}) during 17:42$-$17:46 UT.
The onset time of writhing is coincident with the commencement of the impulsive phase of the associated flare (Figure~\ref{fig10}).
Moreover, the kinematic evolution of the CME is divided into two phases using GCS modeling (Figure~\ref{fig19}(a)). 
The CME speeds up from $\sim$330 km s$^{-1}$ to $\sim$1000 km s$^{-1}$ after the writhing motion. Hence, the prominence eruption is most probably triggered by kink instability.

Using the spectroscopic observations of two tornadoes by the Interface Region Imaging Spectrograph \citep[IRIS;][]{Dep2014} in Mg {\sc ii} k 2796 {\AA} and Si {\sc iv} 1393 {\AA},
\citet{Yang2018} found coherent and stable redshifts and blueshifts across the tornado axes lasting for $\geq$2.5 hr, 
which is explained by rotating cool plasmas with temperatures of 0.01$-$0.1 MK along a relatively stable helical magnetic structure.
\citet{Pan2014} concluded that the apparent tornado-like prominences result either from counterstreaming and oscillations or from the projection on the plane of the sky of plasma motion
instead of a real vortical motion around an axis \citep{Wed2012}.
In our study, the eruptive prominence is not a quasi-static tornado at all. The combination of counterclockwise rotation at the leg and the writhing motion during eruption provides clear evidence 
for untwisting motion of a twisted flux rope as a result of kink instability. The cool plasmas observed in 304 {\AA} are frozen in the field lines, 
rather than streaming downward along the lines (see Figure~\ref{fig4}).

The 3D reconstruction with GCS modeling is successful in tracking the positions and configuration of CMEs, which is very helpful for space weather prediction.
In previous works, the modelings focus on CMEs originating from the frontside observed from Earth \citep{Cheng2013,Cheng2014b}.
However, simultaneous observations of CMEs from multiple viewpoints enable us to perform 3D reconstruction of CMEs originating from the farside.
In our study, the prominence rises from behind the western limb and undergoes a long slow-rise phase during 13:30$-$16:30 UT followed by a fast rise.
The related flare ribbons and post-flare loops are only visible in the EUV images observed by STA/EUVI, and no enhancements in SXR emissions are detected by GOES.
The adoption of $\phi=91^{\circ}$ gives satisfactory fittings (Figures~\ref{fig17} and \ref{fig18}). 
It should be emphasized that the quiescent filament is very long (Figures~\ref{fig8} and \ref{fig9}), so that the determination of $\phi$ has an uncertainty (e.g., $\sim$1$^{\circ}$).
In addition, synthetic WL images of the CME are unavailable since the density parameters ($N_e$, $\sigma_{trailing}$, $\sigma_{leading}$) are not taken into account \citep{The2006}.
In the future, applications of the GCS modeling will be extended by using images of coronagraphs working in WL and UV wavelengths, 
such as the Metis \citep{Ant2020} and Heliospheric Imager \citep[SoloHI;][]{How2020} on board Solar Orbiter \citep{Mul2020}
and the Lyman-alpha Solar Telescope \citep[LST;][]{Li2019} on board the Advanced Space-based Solar Observatory \citep[ASO-S;][]{Gan2019} launched on 9 October 2022.

\section{Summary} \label{s-Sum}
In this paper, we report the multiwavelength observations of an erupting prominence and the associated CME on 13 May 2013. 
The event occurs behind the western limb in the FOV of SDO/AIA.
The main results are summarized as follows:
\begin{enumerate}
\item{The prominence is supported by a highly twisted magnetic flux rope and shows rapid rotation in the counterclockwise direction during the rising motion.
The rotation of the prominence lasts for $\sim$47 minutes. 
The average period, angular speed, and linear speed are $\sim$806 s, $\sim$0.46 rad min$^{-1}$, and $\sim$355 km s$^{-1}$, respectively.
The total twist angle reaches $\sim$7$\pi$, which is considerably larger than the threshold for kink instability.
Writhing motion during 17:42$-$17:46 UT is clearly observed by SWAP in 174 {\AA} and STB/EUVI in 304 {\AA} after reaching an apparent height of $\sim$405\,Mm.
Therefore, the prominence eruption is most probably triggered by kink instability. A pair of conjugate flare ribbons and post flare loops are created and observed by STA/EUVI. 
The onset time of writhing motion is consistent with the commencement of the impulsive phase of the related flare.}
\item{The 3D morphology and positions of the associated CME are derived using the GCS modeling.
The kinetic evolution of the reconstructed CME is divided into a slow-rise phase ($\sim$330 km s$^{-1}$) and a fast-rise phase ($\sim$1005 km s$^{-1}$) by the writhing motion.
The edge-on angular width of the CME is a constant (60$^{\circ}$), while the face-on angular width increases from 96$^{\circ}$ to 114$^{\circ}$, indicating a lateral expansion.
The latitude of the CME source region decreases slightly from $\sim$18$^{\circ}$ to $\sim$13$^{\circ}$, implying an equatorward deflection during propagation.
To the best of our knowledge, this is the first 3D reconstruction of a CME caused by the eruption of a flux rope due to kink instability.
More case studies are required to perform in-depth investigations of their evolutions.}
\end{enumerate}

\begin{acks}
The authors appreciate the referee for valuable and constructive suggestions to improve the quality of this article.
We thank Drs. Xiaoli Yan and Zhike Xue in Yunnan Observatories for their helpful discussions.
SDO is a mission of NASA\rq{}s Living With a Star Program. The AIA data are courtesy of the NASA/SDO science teams.
STEREO/SECCHI data are provided by a consortium of the US, UK, Germany, Belgium, and France.
This work is supported by the National Key R\&D Program of China 2022YFF0503003 (2022YFF0503000), 2021YFA1600500 (2021YFA1600502) and NSFC grants (No. 11790302, 11790300).
\end{acks}


\bibliographystyle{spr-mp-sola}
\bibliography{sola}

\end{article}
\end{document}